\documentclass[12pt]{article}

\newcommand{\cas}{{\mbox{\footnotesize$\cal S$}}}
\newcommand{\stot}{\cas_{\rm tot} }

\newcommand{\trunc}{{\rm tr}}
\newcommand{\eord}{\EE}
\newcommand{\esup}{\EE_{\rm S}}
\newcommand{\esupsup}{\hat\EE_{\rm S}}
\newcommand{\tre}{\check{\EE}_{(d,D)}}
\newcommand{\cqfd}{{\hfill \underline{{\it q.e.d.}}}}

\newcommand{\paral}{Par(d,D)}

\newcommand{\dd}{{d}} 
\newcommand{\trd}{\check{\dd} \,}

\newcommand{\td}{{\tilde d}}
\newcommand{\tO}{{\tilde\Omega}}

\newcommand{\bpsi}{{\Psi}}
\newcommand{\bfi}{{\Phi}}

\newcommand{\hth}{{\lc\hat\theta_r \rc}}
\newcommand{\hthe}[1]{{\lc\hat\theta_{r_{#1}} \rc}}
\newcommand{\hFI}{{\hat\Phi}}

\newcommand{\hHH}{{\hat\HH}}

\newcommand{\trR}{\check{R} \,}
\newcommand{\trO}{\check{\Omega}}
\newcommand{\trH}{\check{H}}
\newcommand{\trHH}{\check{\ZZ}}
\newcommand{\trFI}{\check{\Phi}}
\newcommand{\trP}{\check{P}}

\newcommand{\eqbul}{\bullet\ }

\newcommand{\hO}{{\hat\Omega}}
\newcommand{\hA}{{\hat A}}

\newcommand{\hd}{{\hat d}}
\newcommand{\hF}{{\hat F}}
\newcommand{\psou}{{\underline p}}

\newcommand{\dth}{\pa_\theta}

\newcommand{\tQ}{{\tilde Q}}

\makeatletter
\@addtoreset{equation}{section}
\makeatother

\def\ii{\'\i}

\def\cao{\c c\~ao}

\def\ftoday{{\sl {Le \number\day \space\ifcase\month 
\or janvier\or f\'evrier\or mars\or avril\or mai
\or juin\or juillet\or ao\^ut\or septembre\or octobre
\or novembre \or d\'ecembre\fi\space \number\year}}}    
\def\ptoday{{\sl {\number\day \space de\space \ifcase\month 
\or janeiro\or fevereiro\or mar{\c c}o\or abril\or maio
\or junho\or julho\or agosto\or setembro\or outubro
\or novembro \or dezembro\fi\space de\space \number\year}}}    
\def\gtoday{{\sl {Den \number\day. \ifcase\month 
\or Januar\or Februar\or M\"arz\or April\or Mai
\or Juni\or Juli\or August\or September\or Oktober
\or November \or Dezember\fi\space \number\year}}}    
\def\today{{\sl {\ifcase\month
\or January\or February\or March\or April\or May
\or June\or July\or August\or September\or October
\or November \or December\fi \space\number\day,\space 
                                            \number\year}}}
         \newcommand{\D}{\Delta}
\newcommand{\e}{\varepsilon}

\newcommand{\m}{\mu}

\newcommand{\om}{\omega}         \newcommand{\OM}{\Omega}
\newcommand{\p}{\psi}

\newcommand{\f}{{\phi}}           \newcommand{\F}{{\Phi}}
\newcommand{\vf}{{\varphi}}
\newcommand{\XI}{\XI}
\newcommand{\EE}{{\cal E}}

\newcommand{\HH}{{\cal H}}

\newcommand{\MM}{{\cal M}}

\newcommand{\ZZ}{{\cal Z}}
\newcommand{\es}{\\[3mm]}

\newcommand{\sla}{\raise.15ex\hbox{$/$}\kern -.57em} 
\newcommand{\Sla}{\raise.15ex\hbox{$/$}\kern -.70em}

\newcommand{\lp}{\left(}\newcommand{\rp}{\right)}
\newcommand{\lc}{\left[}\newcommand{\rc}{\right]}

\newcommand{\complex}{{\kern .1em {\raise .47ex
\hbox {$\scriptscriptstyle |$}}
    \kern -.4em {\rm C}}}
\newcommand{\real}{{{\rm I} \kern -.19em {\rm R}}}
\newcommand{\rational}{{\kern .1em {\raise .47ex
\hbox{$\scripscriptstyle |$}}
    \kern -.35em {\rm Q}}}
\renewcommand{\natural}{{\vrule height 1.6ex width
.05em depth 0ex \kern -.35em {\rm N}}}

\newcommand{\tr}{{\rm {Tr} \,}}

\newcommand{\pa}{\partial}

\newcommand{\dfrac}[2]{{\displaystyle{\frac{#1}{#2}}}}
\newcommand{\dsum}[2]{\displaystyle{\sum_{#1}^{#2}}}   
\newcommand{\dint}{\displaystyle{\int}}

\newcommand{\twiddle}{\lower.9ex\rlap{$\kern -.1em\scriptstyle\sim$}}

\newcommand{\equ}[1]{(\ref{#1})}
\newcommand{\eq}{\begin{equation}}
\newcommand{\eqn}[1]{\label{#1}\end{equation}}
\newcommand{\eea}{\end{eqnarray}}
\newcommand{\eqa}{\begin{eqnarray}}
\newcommand{\eqan}[1]{\label{#1}\end{eqnarray}}
\newcommand{\ba}{\begin{array}}
\newcommand{\ea}{\end{array}}
\newcommand{\eqac}{\begin{equation}\begin{array}{rcl}}
\newcommand{\eqacn}[1]{\end{array}\label{#1}\end{equation}}

\begin{document}

%%%%%%%%%%%%%%%%%%%%%%%%%%%%%%%%%%%%%%%%%%%%5
%{topym.tex \hfill \today}
%{tgft.tex \hfill \today}
%\vspace{3mm}
%%%%%%%%%%%%%%%%%%%%%%%%%%%%%%%%%%%%%%%%%%%%%

\hspace*{\fill}{{\normalsize 
\begin{tabular}{l}
%{\sf } \\
{\sf LYCEN 2002-54}  \\
{\sf UFES-DF-OP2003/1}  \\
 {\sf \today} \\
 \end{tabular}   
 }}

\begin{center}
{\LARGE\bf Observables}
\end{center}
\begin{center}
{\LARGE\bf in}
\end{center}
\begin{center}
{\LARGE\bf Topological Yang-Mills Theories}
\end{center}

\vspace{3mm}

\begin{center}{\large 
Jos\'e Luis Boldo$^{*,}$\footnote{Supported in part by the 
{\it Conselho Nacional de
Desenvolvimento Cient\'\i fico e Tecnol\'ogico (CNPq -- Brazil)}.},
Clisthenis P. Constantinidis$^{*,**,1,}$\footnote{Supported in part by the 
{\it Coordena\cao\ de Aperfei\c coamento de Pessoal de N\ii vel Superior
(CAPES -- Brazil)}.}, \es
Fran\c cois Gieres$^{***}$,
Matthieu Lefran\c cois$^{***}$
and Olivier Piguet$^{*,1}$
}\end{center}
\vspace{1mm}

\noindent $^{*}$ {\it Universidade Federal do Esp\'{\i}rito Santo 
(UFES), 
CCE, Departamento de F\'{\i}sica, Campus Universit\'ario
de Goiabeiras, BR-29060-900 - Vit\'oria - ES (Brasil)}
%\vspace{1mm}

\noindent $^{**}$ {\it The Abdus Salam ICTP, Strada Costiera 11, 
I - 34014 - Trieste (Italy)}
%\vspace{1mm}

\noindent $^{***}$ {\it 
Institut de Physique Nucl\'eaire,
Universit\'e Claude Bernard (Lyon 1), \\
43, boulevard du 11 novembre 1918,
      F - 69622 - Villeurbanne (France)}
%\vspace{1mm}

{\tt E-mails: jboldo@cce.ufes.br,
clisthen@cce.ufes.br, \\
gieres@ipnl.in2p3.fr, 
lefrancois@ipnl.in2p3.fr, piguet@cce.ufes.br}

%%%%%%%%%%%%%%%%%%%%%%%%%%%%%%%%%%%%%%%%%%%%%%%%%%%%%%%%%%%%%%%
\vspace{8mm}

{\small 
\noindent
{\bf Abstract:} 
Using topological Yang-Mills theory as example, 
we discuss the definition and  
determination of observables in topological field theories
(of Witten-type) 
within the superspace formulation proposed by Horne. 
This approach to the equivariant cohomology 
leads to a set of bi-descent equations 
involving the BRST and supersymmetry operators 
as well as the exterior derivative. 
This allows us to determine superspace expressions 
for all observables, and thereby to 
recover the Donaldson-Witten polynomials when choosing 
a Wess-Zumino-type gauge. }

\newpage 

\tableofcontents

%%%%%%%%%%%%%%%%%%%%%%%%%%%%%%%%%%%%%%%%%%
\section{Introduction}

Topological field theories have been introduced some fifteen years 
ago~\cite{witten, witten-knots} and continue to represent a field 
of active interest, e.g. see ref. \cite{bau-tanz, bgz, bgeyer}. 
The purpose of the present work is to come back to the issue of 
determining all of the observables for these theories (for some 
general reviews, 
see ref.~\cite{birm}). These  observables are of a global nature, 
e.g. knot invariants in
 Chern-Simons theory~\cite{witten-knots} or Donaldson
invariants in 
 topological Yang-Mills (YM) theory~\cite{witten}
as well as the counterparts of the latter in topological  gravity
\cite{topgrav}. 
For topological YM and gravity theories, these observables 
belong to the so-called equivariant cohomology as originally 
shown by Witten in his
pioneering work on $4d$ topological YM theory~\cite{witten}
and further elucidated in the sequel
from the mathematical point of view~\cite{ouvry}.  
Equivariant
cohomology amounts to computing the cohomology of a 
supersymmetry-like operator 
$\tQ$ (which is the BRST operator associated to the local shift
symmetry of gauge fields) 
in the space of {\em gauge invariant}
local functionals of the fields. 
A crucial point is that the cohomology of $\tQ$, although empty
in the space of the unrestricted local functionals, becomes
nonempty if gauge invariance is imposed on these 
functionals~\cite{ouvry, delduc}.

As pointed out by Horne~\cite{horne}, 
the supersymmetry operator 
may be represented as the derivative with respect to a Grassmann-odd  
parameter $\theta$ within a superfield formalism 
in which gauge invariance is implemented as 
supergauge invariance following the introduction of a superconnection. 
Although  superfield formulations of this type have been found  
to be quite useful for the discussion of the dynamics 
and symmetries of topological models of Witten-type 
(also termed cohomological field theories) 
\cite{horne, ouvry,results}, they have 
not been considered so far for the determination of observables. 
The present paper 
fills this gap and shows that one can directly apply 
the powerful methods and results of the BRST cohomology
associated to (super)gauge invariance~\cite{barnich}. 
This provides a complete basis of observables and 
-- as expected -- it allows us to recover Witten's results which have  
been tackled using other 
approaches in the past~\cite{bau-sin, delduc, blasi, bauer}.

We shall be fairly explicit in our presentation since the 
present work will serve as a basis 
for the systematic study~\cite{clisthen,boldo2} of 
more complex
models involving equivariant cohomology
like topological gravity in various 
dimensions~\cite{topgrav}
and YM theories with more
than one supersymmetry generator~\cite{geyer, bgeyer}.
We note that 
the techniques that we develop for 
the treatment of bi-descent equations 
should also be useful in other contexts where equations of this 
type appear, e.g. see~\cite{bgz}.

Our paper is organized as follows. 
In section \ref{symetries}, 
we present the general 
framework and, in particular, the BRST formalism 
for topological YM theories in the superspace 
associated with the shift supersymmetry. 
If the  supergauge invariance is fixed by a Wess-Zumino type 
condition, we recover the field content and transformation laws 
that have been 
considered in the original literature~\cite{witten,ouvry,bau-sin}.
In section \ref{observables in SF form}, we determine 
the cohomology of the BRST
operator in the functional space constrained by 
the requirements of supersymmetry 
invariance and zero ghost-number.
We shall see that it corresponds to
a certain
 subset of the cohomology  
$H(\cas | d)$ 
 of the BRST operator $\cas$ modulo the exterior derivative $d$ 
in the space of differential forms whose coefficients
are superfields.
 Some explicit examples are presented in section \ref{app-examples}. 
An appendix gathers the proofs of several lemmas and 
propositions presented in the main body of the text. 

Although the formalism is motivated by $4$-dimensional topological
YM theories, the value of the spacetime dimension
will not be specified. 
In fact, we shall not consider the dynamics nor 
address the problem of gauge-fixing
(requiring the introduction of antighosts
and Lagrange multiplier fields)
and thereby our results have a purely algebraic character. 
In particular, they are completely independent of the 
spacetime dimension.

%%%%%%%%%%%%%%%%%%%%%%%%%%%%%%%%%%%%%%%%%%
\section{Symmetries}\label{symetries}     %%%%%%%%Sect. 2
%%%%%%%%%%%%%%%%%%%%%%%%%%%%%%%%%%%%%

Topological field theories of Witten-type 
can be obtained from extended supersymmetric 
gauge theories by performing 
an appropriate twist. The invariance under extended 
supersymmetry transformations
then gives rise to a shift symmetry in the  
topological model. Thus, the latter invariance 
is often referred to as 
supersymmetry transformation and it can be conveniently 
described in a superspace~\cite{horne,ouvry}. 
The superspace formulation that we shall use is the one 
of Horne~\cite{horne}, though the latter author did
not elaborate on supergauge transformations 
whose inclusion is essential for the discussion
of observables. Let us first introduce superspace and 
the geometric objects that it supports. 
   
%%%%%%%%%%%%%%%%%%%%%%%%%%%%%%%%%%%%%%
\subsection{Superspace}\label{superspace}

We extend the $n$-dimensional
spacetime manifold by a single Grassmannian  
variable $\theta$ so as to obtain a superspace parametrized 
by local coordinates $(x, \theta)$. 
We assign a ``supersymmetry-number" 
({\em SUSY-number} or {\em SUSY-charge} for short) 
to all fields 
and variables\footnote{Originally this number 
was referred to as ``ghost-number'' $U$ 
\cite{witten}.}:
for the variable $\theta$, this number is $-1$
and, quite generally, an upper or lower $\theta$-index 
on a field corresponds to a SUSY-number $-1$ or $+1$, respectively. 

A {\em superfield} is a function on superspace, 
\begin{equation}
\label{sf}
F(x,\theta) = f(x) + \theta f_{\theta}'(x)
\, ,
\end{equation}
where $f(x)$ has the same Grassmann parity as $F(x,\theta)$
while its superpartner $f_{\theta}'(x)$ has the opposite parity.
To be more precise, the superfield (and thereby its components) 
is also supposed to transform in a specific way under
supersymmetry transformations, 
see eqs.(\ref{def-Q}) below. 

A {\em $p$-superform} admits the expansion 
\eq
\hO_p (x,\theta) = \dsum{k=0}{p} \, \OM_{p-k}(x,\theta) \; (d\theta)^k
\, ,
\eqn{s-form}
where $\OM_{p-k}$ has $k$ lower
$\theta$-indices that we did not
 spell out. 
The components $\OM_q (x,\theta)$ of the $p$-superform 
(\ref{s-form}) are $q$-forms
whose coefficients are superfields:
\eq
\OM_q (x,\theta) = \frac{1}{q!} \, \OM_{\m_1\cdots\m_q}(x,\theta) 
\; dx^{\m_1}\cdots dx^{\m_q}
= \om_q (x) + \theta \om_{q\theta} ' (x)
\, .
\eqn{gform}
Superspace expressions of the form \equ{gform} will be referred to as 
{\em superfield forms} in the sequel.
In the expansion \equ{gform} and in the following, 
the wedge product symbol is always omitted.
Moreover, we shall adhere to the notational conventions 
used in the previous expressions:
functions or forms on ordinary spacetime are 
denoted by small case letters 
(e.g. $f,   f_{\theta}', \om_q,  \om_{q\theta} ',\dots$), 
superfields or superfield forms by upper case letters
(e.g. $F, \OM_q,\dots$) 
and $p$-superforms with $p \geq 1$
(e.g. $\hO_p$)
by upper case letters with a
``hat''.

The exterior derivative in superspace is defined by
\eq
\hd = d+d\theta \dth \qquad\mbox{with}\quad d = dx^\m \pa_\m \ .
\eqn{s-ext-der}
We have $0= \hd^{\, 2} = d^{\, 2} = (d\theta \dth)^2 
= [d,d\theta \dth]$ where the bracket 
$[\cdot , \cdot]$ denotes the graded 
commutator.

A global, infinitesimal supersymmetry transformation
 is given by a translation of the $\theta$-variable, i.e.
$\theta \to \theta + \e ^{\theta}$. Thus, it is a supercoordinate
transformation generated by the vector field 
$\e ^{\theta} \pa_{\theta} \equiv  \e ^{\theta} Q$
where $Q= \pa_{\theta}$ represents the supersymmetry generator.
The latter operator is nilpotent (i.e. $Q^2=0$)
and it raises the SUSY-number
by one unit. The {\em supersymmetry transformations}
of the superfield (\ref{sf}) and of its component fields read as
\eq\ba{c}
   Q F =  \pa_{\theta} F 
 \es
 Q f = f_{\theta}'
\quad ,\quad 
Q f_{\theta}' =0
\, .
\ea\eqn{def-Q}
Following standard practice, we use the same symbol $Q$ 
to denote the action of the supersymmetry generator $Q$ 
on either component fields or on superfields, superfield forms and
superforms. On each of the latter, $Q$ acts by virtue of the
$\theta$-derivative.
Thus, any superfield (\ref{sf}) or superform \equ{s-form}
has the general form 
\begin{eqnarray}
F (x, \theta )  \!\!\! &=& \!\!\! f(x) + \theta \, (Qf )(x) 
   \nonumber \\
\label{gensu}
\hO_p (x,\theta)  \!\!\! &=& \!\!\! \hO_p (x) + \theta \, (Q \hO_p ) (x)
\qquad {\rm with} \ \; 
\hO_p (x ) = \dsum{k=0}{p} \, \OM_{p-k}(x ) \; (d\theta)^k
\ .
\end{eqnarray}

While an ordinary $p$-form can be integrated over 
a manifold of dimension $p$, there is no 
directly analogous theory of integration 
for superforms \cite{kostant}. 
Yet one can introduce some algebraic integration rules
which are quite useful for the discussion of descent 
equations in the BRST formalism. 
To do so, we consider a collection ${\cal M}= \left(M_0,M_1,\dots,M_p 
\right)$
of closed 
spacetime manifolds $M_k$ of dimension $k$
and we define the spacetime integral of a $p$-superform 
on this collection by
the direct sum\footnote{A manifold of dimension $0$ represents
a point, $M_0 = \{ y \}$, and 
$\int_{M_{0}} \OM_{0}(x,\theta) \equiv  \OM_{0}(y,\theta)$.}
\eq
\int_{\cal M} \hO_p (x,\theta) \equiv \sum_{k=0}^{p} 
(d\theta)^k \int_{M_{p-k}} \OM_{p-k}(x,\theta)
\, .
\eqn{integral}
This expression still depends on $\theta$. 
Since integration with respect to the Grassmannian variable
$\theta$ means derivation with respect to $\theta$ (i.e. the
operation $Q= \pa_{\theta}$), we set
\[
\int_{\theta}\int_{\cal M} \hO_p (x,\theta) 
= Q \int_{\cal M} \hO_p (x,\theta)
\]
or, more explicitly, 
\eq
\int_{\theta}\int_{\cal M} \hO_p (x,\theta) \equiv 
\sum_{k=0}^{p} 
(d\theta)^k \int_{M_{p-k}} Q \OM_{p-k}(x,\theta)
= 
 \sum_{k=0}^{p}  (d\theta)^k 
\int_{M_{p-k}} \om_{p-k}'(x)
\, .
\eqn{s-integral}
The so-defined expression can be referred to as 
superspace integral of a $p$-superform.

%%%%%%%%%%%%%%%%%%%%%%%%%%%%%%%%%%%%%%
\subsection{BRST-formalism}\label{BRST-formalism}
   
Within the BRST-formalism, the parameters of infinitesimal 
symmetry transformations are turned into ghost fields.
The latter have {\em ghost-number} $g=1$ while the 
fundamental fields
appearing in the invariant action 
(i.e. the connection for topological YM theory)
have
a vanishing ghost-number.
 The Grassmann parity of an object is given by the
parity of its {\it total degree} defined as the sum $p+g+s$ of 
its form degree $p$,  its ghost-number $g$ and its SUSY-number $s$.
All commutators and  brackets are assumed to be graded 
according to this grading.

%%%%%%%%%%%%%%%%%%%%%%%%%%%%%%%%%%%%%%%%%%%%
\subsection{Topological Yang-Mills theory in 
superspace}\label{topym in s-space}

The basic variables are the connection $1$-superform 
$\hA(x,\theta)$
and the ghost superfield $C(x,\theta)$ which corresponds  to
infinitesimal supergauge transformations. 
These variables are Lie algebra valued, i.e.  
\[
\hA= \hA^a T_a
\quad ,\quad 
C = C^a T_a
\quad , \quad 
[T_a,T_b] = f_{ab}{}^c T_c
\ ,
\]
where the matrices $T_a$ represent the generators of the
Lie group that is  
chosen as structure group of the theory.

The BRST transformations of $\hA$ and $C$ 
describing the supergauge invariance of the theory 
read as 
\eq
\cas \hA = -( \hd C +[ \hA ,C ] ) 
\quad  ,\quad 
\cas C =  -C^2 
\ .
\eqn{BRS-super}
The so-defined BRST operator $\cas$ is nilpotent,
i.e. $\cas ^2 =0$. 

Let us now introduce the
components of the $1$-superform
$\hA$, 
\eq
\hA = A(x,\theta) + d\theta  \, A_\theta(x,\theta)\ ,
\eqn{agf}
as well as the spacetime components of all superfield forms:
\begin{eqnarray} 
A (x,\theta)  \!\!\! &=& \!\!\!   a(x)+\theta \, \psi_{\theta} (x) 
\quad , \quad 
C (x,\theta)  =  c(x) + \theta \, c'_{\theta} (x)
\nonumber \\
A_\theta (x,\theta)  \!\!\!&=&\!\!\!   \chi_{\theta} (x)
+ \theta \, \f_{\theta \theta} (x) 
\ .
\label{aac}
\end{eqnarray}
Here, $a$ denotes the connection $1$-form associated 
to ordinary gauge transformations and $c$ the corresponding ghost.
In the sequel, 
the covariant derivative with respect to $a$
will be denoted by ${D_a}c   = dc + [a,c]$ 
and the $\theta$-indices labeling spacetime fields
will 
 be omitted in order to simplify the notation.

Substitution of \equ{agf} into \equ{BRS-super} yields 
the BRST transformations of $A$ and $A_\theta$, 
\eq
\cas A= -(dC +[A,C]) \equiv -{D_A} C\ ,\quad 
\cas A_\theta =  
- (\pa_\theta C + [A_\theta, C ]) 
\eqn{s-superfields}
and the expansions \equ{aac} provide 
the {\em BRST transformations of the spacetime fields:} 
  \eq\ba{ll}
\cas a =  -{D_a}c  \ ,\qquad &\cas c  =  -c^2 \es 
\cas\p = - [c,\p]  - D_a c'  
\ ,\qquad &  \cas c' =  -[c,c'] \es
\cas\f = - [c,\f] - [\chi,c'] \ ,\qquad &   \cas\chi =   - [c,\chi] - c'  
\ .
\ea\eqn{BRS-comp} 
The {\em supersymmetry transformations} of all component
fields appearing in \equ{aac}
follow from \equ{def-Q}:
  \eq\ba{lll}
 Qa =  \p  \ ,\qquad & Q\chi =  \f  \ ,\qquad & Qc = c' 
 \es 
 Q\p =0 \ ,\qquad & Q\f =0 \ ,\qquad & Qc' =0  
\ .
\ea\eqn{ssc} 
We have the graded commutation relations 
\[
[ \cas , Q] = [ \cas , d] = [d,Q] = 0 
\ . 
\]
It is quite useful to consider the following redefinitions
of superfields: 
\eq\ba{rl}
\bpsi \!\!\! & \equiv \partial_{\theta} A + D_A A_\theta 
= \psi +  D_A A_\theta \es
\bfi  \!\!\!& \equiv  \partial_{\theta} A_\theta + A_\theta ^2 
= \f+ A_\theta^2  \es 
K  \!\!\!& \equiv  -( \pa_\theta C + [A_\theta,C] )  
= -c' - [A_\theta, C] \ .
\ea\eqn{s-field-red}
In fact, in terms of these expressions, 
the $d\theta$-expansion of the supercurvature form 
$\hF=\hd\hA +\hA^2$ reads as 
\eq
\hF = {F_A} + \bpsi \, d\theta  + \bfi \, (d\theta)^2 \ , 
\quad \mbox{with} \quad {F_A} = dA+A^2 \ , 
\eqn{s-curvature}
while the BRST transformations read as 
  \eq\ba{ll}
\cas A =  -{D_A}C  \ ,\qquad & \cas C  =  -C^2 \es 
\cas\bpsi = - [C,\bpsi] \ ,\qquad &  \cas\bfi = - [C,\bfi] \es
\cas A_\theta =  K \ , \qquad &   \cas K = 0 \ ,  
\ea\eqn{BRS-redef} 
and the supersymmetry transformations are given by 
\eq\ba{rll}
QA \!\!\! & =  \bpsi - {D_A} A_\theta \ , 
& \qquad Q \bpsi  = - {D_A} \bfi - [ A_\theta, \bpsi ] \es 
Q{F_A} \!\!\! &= -{D_A}\bpsi - [A_\theta,{F_A}] \ , 
& \qquad  Q \bfi  = -[A_\theta,\bfi] \ .
\ea\eqn{susy-redef} 
We note that $Q$ acts on $A,\bpsi ,\bfi$ and the curvature  
$F_A$ according to 
\[
Q= Q_0 + \left. \cas \right| _{C= A_\theta} \ ,
\]
with 
\begin{eqnarray}
\label{qqq}
Q_0 A  \!\!\! &= & \!\!\!      \bpsi  \ , \quad
Q_0 \bpsi = - {D_A}\bfi \ , \quad
Q_0 \bfi =0 \\
(Q_0)^2  \!\!\! &= & \!\!\!  \mbox{infinitesimal
supergauge transformation with parameter} \ \bfi \ .
\nonumber
\end{eqnarray}
Thus, the operator $Q_0$ is nilpotent when acting on an invariant
polynomial depending on the variables $F_A, \bpsi ,\bfi,
D_A\bpsi ,D_A\bfi$. 

In this paper, all space-time forms will be taken as polynomials of the
basic forms $a,\,\psi,\,\chi,\,\f,\,c,\,c'$ and their $d$-derivatives.
Superfield forms and superforms will be taken as polynomials of the
basic superfield forms $A,\,A_\theta,\,C$ and their $\dth$- and 
$d$-derivatives. 
Since we only discuss the kinematics, we do not fix {\em a priori}
the space-time dimension.
The respective functional spaces will be denoted by
\eq
\eord \, :\  \mbox{space-time forms}\ ,\qquad
\esup \, : \ \mbox{superfield forms}\ ,\qquad
\esupsup \, : \ \mbox{superforms}\ .
\eqn{funct-spaces}

We conclude this section with two results 
which will be important for our investigations:
%%%%%%%%%%%%%%%%%%%%%%%%%%%%%%%%%%%%%%%%%
\begin{enumerate}\item[]
{\bf Proposition \ref{symetries}.1}
The cohomology $H(Q)$ of the supersymmetry operator $Q$ in the space
$\eord$, $\esup$ or $\esupsup$ is trivial, i.e. 
\eq\ba{l}
\mbox{If} \ \; Q \varphi =0 \ ,\quad
\mbox{then}\quad \varphi = Q \varphi^{\prime} \ , \es 
\mbox{with both} \  \varphi \ \mbox{and} \ \varphi^{\prime}
\ \mbox{belonging to either} \ \eord\,,\ \esup\,\ \mbox{or }\esupsup 
\ . 
\ea\eqn{prop2-1}
\end{enumerate}
\noindent{\bf Proof:} 
For the functional space $\eord$,the proof follows from the fact 
that all fields represent $Q$-doublets $\{f,f'\}$ with $Qf=f'$ and
$Qf'=0$, and from a well-known result according to which such
doublets do not contribute  to the cohomology 
(e.g. see proposition 5.8. of reference~\cite{pig-sor}). 
The extension of this result to the spaces
$\esup$ and $\esupsup$ is straightforward, 
the action of the operator 
$Q$ being given on these spaces by 
the derivative $\dth$. \cqfd 
%%%%%%%%%%%%%%%%%%%%%%%%%%%%%%%%%%%%%%%%%%%%%%%%%%
\begin{enumerate}\item[]
{\bf Proposition \ref{symetries}.2} (``Algebraic Poincar\'e Lemma'').
The cohomology $H(d)$ of the exterior derivative $d$ in the
space $\eord$, $\esup$ or $\esupsup$ is trivial.
\end{enumerate}
\noindent{\bf Proof:} The result for the space $\eord$ 
is well known~\cite{barnich} within the present context 
where the space-time dimension is not fixed a priori.  
The extension to the spaces $\esup$ or $\esupsup$ follows 
by considering an expansion in $\theta$ or in $\theta$ and $d\theta$,
respectively, 
and by using the linearity of $d$. \cqfd

%%%%%%%%%%%%%%%%%%%%%%%%%%%%%%%%%%%%%%%%%%%
\subsection{Topological Yang-Mills in the Wess-Zumino 
gauge}\label{tym in WZ-gauge}

\subsubsection{Symmetry transformations in the 
WZ-gauge}\label{transf-wz-gauge}
   
The supergauge freedom can be reduced to the ordinary gauge freedom
by imposing the 
{\em Wess-Zumino (WZ) supergauge} condition
\eq
\chi = 0\ .
\eqn{wz-cond}
By virtue of eqs.\equ{BRS-comp}, the $\cas$-invariance of this
condition
 requires $c' =0$. The $\cas$-variations \equ{BRS-comp} 
then reduce to the
{\em BRST transformations in the WZ-gauge} which describe  
ordinary gauge transformations:
\eq
\cas a=-D_a c \quad ,\quad 
\cas\p=-[c,\p]
\quad ,\quad 
\cas\f=-[c,\f]
\quad ,\quad 
\cas c=-c^2\ .
\eqn{BRST-WZ}
Condition \equ{wz-cond}
is not invariant under the
SUSY generator $Q$, i.e. under the variations \equ{ssc}.    
A modified
SUSY operator $\tQ$ which leaves this condition 
stable is obtained by
combining $Q$ with a compensating BRST transformation (\ref{BRS-comp}) 
according to
 \eq
\tQ = Q + \left. \cas \right| _{\chi = c=0,\, c'=\f}
\qquad 
{\rm on} \ \; a, \p, \f  
\ .
\eqn{WZ-Q}
Thus, we get the {\em supersymmetry transformations in the 
WZ-gauge}, 
\eq\ba{ll}
\tQ a = \p
\quad ,\quad 
\tQ \p =  
-D_a\f
\quad  ,\quad 
\tQ \f = 0
\ ,
\ea\eqn{Q-transf-WZ}
which satisfy
\eq
\tQ^2 = \mbox{infinitesimal
gauge transformation with parameter} 
\ \f\ .
\eqn{nilp} 
A crucial point of the theory is the fact that the operator 
$\tQ$ is nilpotent when acting on invariant polynomials, 
very much like the operator $Q_0$ defined in \equ{qqq}. 
We also note that the algebra generated 
by the forms $a$, $\psi$ and $\f$
and their exterior derivatives,  
together with the action of the operators  $\cas$ and $\tQ$ 
given by \equ{BRST-WZ} and \equ{Q-transf-WZ} 
is isomorphic to the algebra generated by the superfield forms
$A$, $\bpsi$ and $\bfi$ (defined in eq.\equ{s-field-red})
and their exterior derivatives,  
together with the action of the 
operators  $\cas$ and $Q_0$ given by 
\equ{BRS-redef} and \equ{qqq}. 

The supersymmetry-BRST formalism defined by 
eqs.\equ{Q-transf-WZ} and \equ{BRST-WZ}
is the one used by Witten in his pioneering  
work on four-dimensional topological
YM theory~\cite{witten}. 
We emphasize that the only ghost field in this approach is $c$
(as well as $c'$ in a general supergauge). This fact is in 
contrast to some other approaches where $\p$ and $\f$ have 
ghost-numbers $1$ and $2$, respectively (e.g. 
see~\cite{ouvry, bau-sin})\footnote{In fact, in these approaches 
our ghost- and SUSY-numbers are added together so as to yield a single 
BRST ghost-number.}.

For later reference, we display the $\theta$-expansion
of the superconnection $\hA$ and of the associated 
curvature $\hat F = \hd \hA + \hA ^2$ in the WZ-gauge 
(cf. \equ{s-curvature}):
\begin{eqnarray}
\hA  \!\!\! &=& \!\!\!  
a + \theta \, \psi + \theta  d\theta \, \f
   \nonumber \\
   \label{wzf} 
   \hat F \!\!\! &=& \!\!\!  {F_a} + \psi \, d\theta + \f \, (d\theta)^2 
  - \theta \, {D_a} \psi - \theta  d\theta \, {D_a}\f    
\ . 
\end{eqnarray}

%%%%%%%%%%%%%%%%%%%%%%%%%%%%%%%%%%%%%%%%%%%%%%%%%%%%%%%
\subsubsection{Witten's observables and descent 
equations}\label{witten desc eq}

The expression \equ{wzf} of $\hat F$ has the form 
\eq
\hat F = {\cal F} + \theta \, \tQ {\cal F} 
\qquad {\rm with} \ \;  
{\cal F} \equiv   {F_a} + \psi \, d\theta + \f \, (d\theta)^2 
\, ,
\eqn{univ}
i.e. it is of the same form as a generic superform 
in a general gauge, cf. eq.\equ{gensu}.
More specifically, one can check that we have  
\eq
\tQ {\cal F} =
- (D_a {\cal F}) 
\, (d\theta)^{-1}
\, , 
\eqn{aqt} 
where the notation $(d\theta)^{-1}$ is symbolic, though 
it can be further justified. 

The quantity ${\cal F}$ represents the curvature 
of the universal bundle considered by 
Baulieu and Singer~\cite{bau-sin}
in their derivation of Witten's observables. 
(Actually, these authors 
did not introduce the monomials $d\theta$, rather they 
associated ghost-numbers $1$ and $2$ to  
$\psi$ and $\f$, respectively.) 
For the derivation of observables, we can argue as follows.
For $m=1,2,\dots$, we have 
\[
\tr \hat F ^m = \tr {\cal F} ^m + \theta \, \tQ  \, \tr {\cal F} ^m
\, , 
\]
where the first term yields the 
{\em Donaldson-Witten polynomials},
\begin{eqnarray}
\tr {\cal F} ^m   \!\!\! &=& \!\!\!  \tr {F_a}^m 
 + \tr (m {F_a}^{m-1} \psi ) d\theta +  \cdots 
  +  \tr (m \psi \f ^{m-1}  )  (d\theta)^{2m-1} 
  + \tr \f^m  (d\theta)^{2m}
\nonumber \\
 \!\!\! & \equiv & \!\!\!  
 \dsum{p=0}{2m}  \; \omega_p \; (d\theta)^{2m-p}
\label{expsi}
\end{eqnarray}
 and where the second term represents a total derivative by virtue of 
 eq.\equ{aqt}:
 \[
 \tQ  \, \tr {\cal F} ^m = - d  \, \tr {\cal F} ^m \;  (d\theta)^{-1}
\, .
\]
By substituting the expansion \equ{expsi} into the last relation, 
we obtain {\em Witten's descent equations} for the polynomials $\omega_p$:
\eq
\tQ  \omega_p + d\omega_{p-1} = 0
 \qquad  
(\, p = 0, 1, \dots, 2m \, )  
\ .
\eqn{wdes}
Here and in the following, the forms of negative form degree 
are assumed to vanish by convention.

%%%%%%%%%%%%%%%%%%%%%%%%%%%%%%%%%%%
\subsubsection{Combining all symmetries}\label{all symm}

It is possible to incorporate the transformations 
\equ{Q-transf-WZ}
into the BRST algebra
by introducing a constant
{commuting} ghost $\e$:
the BRST operator then acts on $a, \psi, \f$ according to  
\eq
\stot = \cas +\e \tilde Q
\eqn{s-WZ}
and on $c,\e $ according to 
\eq
\stot c = -c^2 +  \e^2 \f 
\quad , \quad 
\stot \e =0 
\ ,
\eqn{s-c-WZ}
which ensures the nilpotency of the operator $\stot$.
More explicitly, we have the expansion  \cite{delduc}
\eq
\stot = \cas_0 +\e \cas_1 + \e ^2 \cas_2
\ ,
\eqn{exps}
with
\eq\ba{lll}
\cas_0 a =  
-D_a c  
\ ,\qquad &\cas_1 a = \p  \ ,\qquad &\cas_2 a =0 
\es
\cas_0 \p =  - [c,\p]  \ ,\qquad &\cas_1 \p = 
-D_a \f 
\ ,\qquad &\cas_2 \p =0 
\es
\cas_0 \f =  - [c,\f] \ ,\qquad &\cas_1 \f = 0\ ,\qquad &\cas_2 \f =0 
\es
\cas_0 c  =  -c^2  \ ,\qquad &\cas_1 c = 0\ ,\qquad &\cas_2 c =\f
\ .
\ea\eqn{mag} 
where
\eq
\qquad (\cas_0)^2 =0 \ ,\qquad [\cas_0, \cas_1]=0  \ ,\qquad 
(\cas_1)^2+ [\cas_0, \cas_2]=0  
\ .
\eqn{fil}
In terms of the notation introduced above, we
have $\cas_0 = \cas$ and $\cas_1 = \tQ$ on $a,\p, \f$.
If we only consider functionals $\Delta$ depending on $a,\p, \f$
and {\em not} on $c$ (i.e. functionals of zero ghost-number), then the last
relation of \equ{fil} is nothing but \equ{nilp}.
If these functionals are, in addition, gauge invariant, 
then the operator  $\cas_1$ is nilpotent:
\eq
\cas_0 \Delta =  0 = \cas_2 \Delta \quad 
\Longrightarrow \quad (\cas_1)^2 \Delta =0
\ .
\eqn{crux}
Its cohomology is referred to as equivariant cohomology 
and will be further discussed in the next section.
(Thus, equivariant cohomology  
is the cohomology of the operator 
$\cas_1$ in the space of local functionals of 
$a, \p,  \f$ and $c$ restricted  
by $\cas_2$ - and $\cas_0$ - invariance.)

To conclude, we note that the algebra (\ref{s-WZ})(\ref{s-c-WZ}) can
also be obtained along a slightly different, though equivalent line of
reasoning.  In fact, we could include the supersymmetry variations 
generated by
$\e Q$ 
 right away into the BRST transformations (\ref{BRS-comp}):  
the stability of the WZ-condition (\ref{wz-cond})
then restricts the ghost field $c'$ to be equal to
$\e \f$ and readily yields the results (\ref{s-WZ})(\ref{s-c-WZ}).
The decoupling \equ{BRST-WZ}\equ{Q-transf-WZ} is 
realized \cite{delduc}
by considering the filtration $N= \e \, \partial /\partial \e$ 
and  the expansion \equ{exps}.

%%%%%%%%%%%%%%%%%%%%%%%%%%%%%%%%%%%
\section{Observables in the superspace 
formalism}\label{observables in SF form}
%%%%%%%%%%%%%%%%%%%%%%%%%%%%%%%%%

In the following, the expression 
$^s  \vf^g_p$ denotes a 
$p$-form $\vf$ of ghost-number $g$ and SUSY-number $s$.

%%%%%%%%%%%%%%%%%%%%%%%%%%%%%%%%%%%%
\subsection{Equivariant cohomology and 
Witten's observables}\label{equiv coho witten obs}

Let us first consider the {\em WZ-gauge} setting described in
the preceding
 section since the latter has been chosen in 
all former discussions of observables.
The representation (\ref{exps})-(\ref{fil}) of the complete 
set of symmetry transformations is quite useful for specifying the 
cohomological characterization of observables.
As is well known, the cohomology of the operator $\stot$
is  empty~\cite{ouvry}. 
Not so the {\it equivariant
cohomology}
which can be described in several 
different 
ways~\cite{witten,ouvry}.
As mentioned in the last section, it can be characterized 
as the cohomology of the operator $\tQ$ 
(defined by \equ{Q-transf-WZ}) in the space
of the
gauge invariant local functionals of $a, \p, \f$ \cite{witten}.
Thus, {\em at form degree zero}, one looks for a local functional 
$^s \D_{(0)}=\int_{M_0} {}^s  \om_0^0 (x)$
which solves the $\tQ$-cocycle condition, i.e. 
\eq
\tilde Q \;{}^s  \D_{(0)} = 0
\ , 
\eqn{Q-cohom} 
and which is constrained by gauge invariance, i.e. 
\eq
\cas \;{}^s  \D_{(0)} = 0
\ . 
\eqn{gconstr}
This cocycle is required to be non-trivial, i.e. 
\eq
{}^s  \D_{(0)} \not =
\tQ \;{}^{s-1}  \D_{(0)}'
\qquad {\rm with} \quad 
\cas \;{}^{s-1}  \D_{(0)}' = 0
\ , 
\eqn{qnontriv}
where $^{s-1}  \D_{(0)}' =\int_{M_0} {}^{s-1}  \om _0^{'0} (x) $.
From the $\tQ$-transformation laws \equ{Q-transf-WZ}, it follows that 
zero-forms cannot be written as a $\tQ$-variation. 
Thus, the non-triviality condition \equ{qnontriv} is automatically satisfied  
at form degree zero. (Note that this is no longer true 
at higher form degree: an expression of the form 
\eq
\int \tQ P_{\rm inv} (F_a, \psi, \phi, , D_a  \psi, D_a \phi)
\ , 
\eqn{mgs}
where $P_{\rm inv}$ is a $\cas$-invariant polynomial, 
is $\tQ$- and $\cas$-invariant, but  $\tQ$-trivial.)
As pointed out by Witten~\cite{witten}, the equations for the 
integrand of ${}^s  \D_{(0)}$, i.e. 
$\tQ \; {}^s  \om_0^0  = 0 = \cas \; {}^s \om_0^0 $, are solved
by the gauge invariant
polynomials $P(\f)$. Thereby, the equivariant 
cohomology is given by the differential forms generated from
these polynomials by virtue of the descent equations
of $\tQ$ modulo $d$, i.e. eqs.\equ{wdes}.
After integrating each of these forms over closed cycles, one obtains
global observables which only depend on the homology class of these
cycles. These observables will be referred to as {\em Witten's observables}.

Equivalently, the equivariant
cohomology can be defined as the
cohomology of the BRST 
 operator $\stot$ restricted to the space of
local functionals
 of  $a, \p,  \f, c$ which are independent of $c$ and 
gauge invariant~\cite{ouvry}.
The mathematical techniques of equivariant cohomology 
\cite{guillemin, bauer} then allow to construct 
some cohomology representatives which turn out 
to coincide with Witten's observables. However, a complete 
determination of the cohomology classes along these lines 
seems to be a difficult task. 

Yet, one can also apply standard cohomological techniques
while working in a restricted functional space.
Using this approach, the authors of reference~\cite{delduc} found that 
the solution of the cohomological problem 
is given by certain $\cas$-cohomology classes 
of ghost-number zero (reproducing again Witten's observables). 
This result suggests to 
look for representatives of the equivariant cohomology within 
the cohomology of the operator $\cas$ (describing gauge
transformations, see (\ref{exps})-(\ref{mag})) in the space of  
local functionals
of  $a, \p,  \f, c$ which are of ghost-number zero and 
invariant under the supersymmetry operator $\tQ$.
From this view-point, one looks for a non-trivial 
solution of the $\cas$-cocycle
condition $\cas \; {}^s  \D_{(d)} =0$
which satisfies the constraint
$\tQ \; {}^s  \D_{(d)} =0$, 
where the non-triviality requirement now 
concerns the $\cas$-operator, i.e. 
\eq
{}^s  \D_{(d)} \not =
\cas \;{}^s  \D_{(d)}'
\qquad {\rm with} \quad 
\tQ \;{}^s  \D_{(d)}' = 0
\ .
\eqn{snontriv}
 However, $\D_{(d)}$ is of ghost-number zero and we do not have any
  fields of ghost-number minus one, therefore the non-triviality condition
  \equ{snontriv} is automatically satisfied
  whatever the form degree $d$.
Thus, at form degree zero, this approach also reduces to 
the cohomology problem (\ref{Q-cohom})(\ref{gconstr})
without any further requirements.
At higher form degree, it regards as {\em non-}trivial 
the solutions of the form (\ref{mgs}) which are trivial 
representatives of equivariant cohomology. 

The latter approach can 
easily 
be extended beyond the WZ-gauge: 
{\em in a general supergauge, 
the  equivariant cohomology
can be determined by looking for 
the ghost-number zero
cohomology classes of the
BRST operator \equ{BRS-super} 
or \equ{BRS-comp} in 
the space of the supersymmetric local functionals} 
(the supersymmetry transformations being 
defined by means of the operator $Q$ according to \equ{ssc}). 
In the following, we shall completely determine this cohomology
while working within the superspace
formalism, only specifying to the WZ-gauge ($\chi =0$)
towards the end.

Thus, let us consider a fixed SUSY-number $s \geq 0$ 
and a fixed degree $d  \geq 0$. 
The task is to find a solution of the {\em cocycle condition}
\eq
\cas\;{}^s \D_{(d)} = 0  \ , 
 \eqn{s-cohomol}
satisfying the {\em SUSY constraint}
\eq
Q\;{}^s  \D_{(d)} = 0
\ . 
\eqn{Q-constraints}
Here, 
\eq 
^s \D_{(d)}=\int_{M_d} {}^s  \om_d^0 (x)
\eqn{observ} 
denotes a local functional
of SUSY-number $s$
which depends on the components of the 
 superfield forms $A, A_{\theta}, C$ and their exterior
derivatives.
Since the solution of 
the problem  \equ{s-cohomol}\equ{Q-constraints}
proceeds in several steps, 
we shall present a summary of results at the end 
of each of the following sections.

Our discussion will be purely algebraic and does not assume a 
specification of the spacetime dimension $n$. 
If the latter is specified, all forms of degree greater than
$n$ vanish identically. 
Those of degree $d$ smaller than  $n$ can be integrated over  
oriented submanifolds $M_d$. The latter manifolds 
are assumed to be closed
which implies the absence of boundary terms upon
integration over $M_d$. 
Thus, we exclude from our discussion the ``trivial'' solution of 
\equ{s-cohomol}\equ{Q-constraints}
which exists for $d=2m$, 
\eq 
^0 \D_{(2m)}= \int_{M_{2m}} {}^0  \om_{2m} ^0 
\equiv
 \int_{M_{2m}} \tr   {F_a}^m \qquad (\, m = 1,2, \dots \, ) \ , 
\eqn{trivobserv} 
since the Pontrjagin density 
$\tr {F_a}^m$ is 
locally given by the exterior derivative of the 
Chern-Simons form of degree $2m-1$. 

Before tackling the cohomological problem in full generality, we already 
note that the determination of observables 
that we presented for the WZ-gauge 
in subsection \ref{witten desc eq} can be generalized to a general
supergauge as follows.

According to equation \equ{gensu},
the curvature $2$-superform has the 
general form
\[
\hF  (x,\theta)  = \hF (x) + \theta \, (Q \hF ) (x)
\ , 
\]
where the first term 
of this expansion
can also be written as
$\left. \hF \right|_{\theta =0} \equiv \hF| $. (In the WZ-gauge, the latter 
expression reduces to the form ${\cal F}$ introduced in eq.(\ref{univ}).)
For $m=1,2,\dots$, 
the $2m$-superform $\tr \hF^m  (x,\theta)$ 
admits an analogous expression: 
\begin{eqnarray}
\tr \hF^m  \!\!\! &=& \!\!\! \tr \hF | ^m 
+ \theta \, Q \tr \hF^m 
\nonumber 
\\
{\rm with} \quad 
\tr \hF |^m  \!\!\! &=& \!\!\! 
\dsum{p=0}{2m} \; ^{2m-p} w^0_{p} \; (d\theta)^{2m-p} 
\ .
\label{genga}
\end{eqnarray}
Since $ \tr \hF^m$ is a closed superform, 
\[
0 = \hd \, \tr \hF^m = (d\theta \partial_{\theta} + d) \, \tr \hF^m 
= (Q \, \tr \hF^m ) \, d\theta + d \, \tr \hF^m 
\ ,
\]
it follows by projection onto the $\theta =0$ component that 
\[
Q \, \tr \hF |^m  
= - (d \, \tr \hF |^m  ) \, (d\theta)^{-1}
\ .
\]
By substituting the expansion (\ref{genga}) into this relation, 
we get {\em Witten's descent equations in a general supergauge}: 
\begin{equation}
Q \;^{2m-p} w^0 _{p } + d \;^{2m-p+1} w^0 _{p -1} = 0 
\qquad (\, p = 0, 1, \dots , 2m \, ) \ .
\label{wdg}
\end{equation}
Explicit expressions for the polynomials $w_p$ 
for $m=1$ and $m=2$ will be given in section \ref{app-examples} 
below and here we only note that 
$\;^{0} w^0 _{2m} = \tr {F_a}^m 
$ whatever the value of $m$.  
The task of the next subsections  is to 
determine if other solutions can 
be obtained by virtue of a systematic study in superspace.

%%%%%%%%%%%%%%%%%%%%%%%%%%%%%%%%%%%%%%%%%%%   
\subsection{The bi-descent equations}\label{bi descent}

In this section, we shall show that the cohomological problem
 \equ{s-cohomol}\equ{Q-constraints}
leads to a set of bi-descent
equations involving superfield forms.
Let us first solve the SUSY constraint \equ{Q-constraints} for
$^s  \D_{(d)}$ given by \equ{observ}.
For the integrand ${}^s  \om_d^0 (x)$, 
it implies
\eq
Q\;^s  \om_d^0 + d\;^{s+1}  \om_{d-1}^0 = 0\ .
\eqn{cond1}
In view of this relation, we shall prove  
the following  proposition:
\begin{enumerate}\item[] 
{\bf Proposition \ref{observables in SF form}.1}   
Let $p$ and $s$ be non-negative integers.  
(Here, we do not refer to the ghost-number which 
only represents a passive label in this proposition.)

(i) The cocycle condition
\eq
Q\;^s  \om_p  + d\;^{s+1}  \om_{p-1}  = 0
\eqn{lemma1a}
implies the  $Q$ modulo $d$ triviality of the space-time form  
$^s  \omega_p $
and the $d$ modulo $Q$ triviality of the space-time form 
$^{s+1} \omega_{p-1} $:
\begin{eqnarray}
^s  \om_p  \!\!\! &=& \!\!\! Q\;^{s-1}  \vf_p  + d\;^{s}  \vf_{p-1} 
\label{lemma1b}\es
^{s+1}  \om_{p-1}  \!\!\! &=& \!\!\! Q\;^{s}  \vf_{p-1}  + d\;^{s+1}  \vf_{p-2} \ ,
\label{lemma1b'}
\end{eqnarray}
with the {\em same} space-time form $^{s}  \vf_{p-1} $ appearing in both
equations.
 
(ii) The same result holds for superfield forms, i.e.  
\eq
Q \;^s  \OM _p  + d\;^{s+1}  \OM_{p-1}  = 0
\eqn{L1a} 
implies
\begin{eqnarray}
^s  \OM_p  \!\!\! &=& \!\!\! Q\;^{s-1}  \F_p  + d\;^{s}  \F_{p-1} 
\label{L1b}\es
^{s+1}  \OM_{p-1}  \!\!\! &=& \!\!\! Q\;^{s}  \F_{p-1}  + d\;^{s+1} 
\F_{p-2} \ . 
\label{L1b'}
\end{eqnarray}
\end{enumerate}
%%%%%%%%%%%%%%%%%%%%%%%%%%%%%%%%%
\noindent{\bf Proof:} See appendix \ref{proof3-1}.

With the help of this proposition, we deduce from \equ{cond1} that
\eq
^s  \om_d^0 =  Q\;^{s-1}  \om_d^0 
\ .
\eqn{Q-omega}
Here and in the following, the total derivative term is suppressed
without loss of generality, since it 
does not contribute to the integrated cocycle $\;^s  \D_{(d)}$.
Furthermore, without loss of generality, we can assume $\;^{s-1} 
\om_d^0$ 
 to be a superfield form 
\eq
^{s-1}  \OM_d^0 = \;^{s-1}  \om_d^0 + \theta \, Q\;^{s-1} \om_d^0  \ ,
\eqn{Q-superfield}
 so that \equ{Q-omega}
reads as
\[
^s  \om_d^0 =  Q\;^{s-1}  \OM_d^0 
\ .
\]
Since the operator $Q$ acts on superfield forms by the  
$\theta$-derivative,
this shows that 
$\;^s\D_{(d)}$ is the superspace integral
of 
$^{s-1}  \OM_d^0$: 
\eq
Q \; ^s  \D_{(d)} = 0 \qquad \Longrightarrow \qquad
^s  \D_{(d)} = \dint_{M_d}Q\;^{s-1}  \OM_d^0 \ .
\eqn{obs-d-s1}

Next, we turn to the cocycle condition \equ{s-cohomol}. 
Since the cohomology of $d$ in the space of local 
field polynomials is trivial~\cite{barnich},
this condition implies 
the {\em descent equations}\footnote{Every form of negative form degree, 
ghost-number or SUSY-number is assumed to vanish by convention.}
\eq 
\cas\;{}^{s}  \om_p^{d-p} + d\;{}^s  \om_{p-1}^{d-p+1} = 0
\qquad
(\, p= 
0,\dots, d \, ) \ .
\eqn{desc-s-d}
\begin{enumerate}\item[]
{\bf Lemma 1}
The SUSY constraint implies that
every form in \equ{desc-s-d}
(and not just the one of highest form degree, i.e. $^{s} \om_d^{0}$) 
can be written as
a SUSY variation, 
\eq
^{s}  \om_p^{d-p} = Q\; ^{s-1}  \OM_p^{d-p} 
\qquad (\, p= 0,\dots, d \, ) \ ,
\eqn{th-deriv}
where $\;^{s-1}  \OM_p^{d-p}$ is a   
superfield form.
\end{enumerate}

%%%%%%%%%%%%%%%%%%%%%%%%%%%%%%%%%%%%%%%%%%%%%%%%%
The proof of this statement proceeds by induction, 
see appendix \ref{proof3-2}. 
%%%%%%%%%%%%%%%%%%%%%%%%%%%%%%%%%%%%%%%%%%%%%%%%%

We note that in eqs.\equ{desc-s-d} and thus in \equ{th-deriv} 
and in the equations to follow, 
the array of descent equations may terminate at some positive 
form degree that we denote by 
$\psou>0$. 
(A simple illustration of such a ``termination of descent''
within abelian gauge theory 
(with field strength $F  = da$)
is given by the cocycles
$\omega_3^0 = a F $ and  $\omega_2^1 = c F$, 
which satisfy $\cas \omega_3^0+ d\omega_2^1 =0$ and 
$\cas \omega_2^1 =0$, so that $\psou =2$
in this case.) 
In such a case, we use the convention that every form 
of form degree less than $\psou$ is vanishing.

%%%%%%%%%%%%%%%%%%%%%%%%%%%%%%%%%%%%%%%%
We are now going to prove:   
\begin{enumerate}\item[]
{\bf Proposition \ref{observables in SF form}.3} 
For given values of $s$ and $d$,
the descent equations \equ{desc-s-d} together with the 
SUSY constraint
\equ{th-deriv} imply a set
of descent equations 
involving superfield forms and all of the three indices:
\eq\ba{c}
\cas\;^{s-r-1}  \OM_{p}^{d-p+r} +  d\;^{s-r-1}  \OM_{p-1}^{d-p+r+1}
+ Q\;^{s-r-2}  \OM_{p}^{d-p+r+1} = 0 \es
(\, r=0,\dots,s-1 \quad ;\quad  p= 0,\dots,d \, ) \ .
\ea\eqn{bi-descent}
We shall call this set of equations 
the {\em bi-descent equations for the pair} $(d,s)$.
\end{enumerate}

\noindent {\bf Proof:} 
In order to derive this result, we first rewrite \equ{desc-s-d}, using
the 
 result
\equ{th-deriv}, as 
\[
Q \lp \cas \;^{s-1}  \OM_{p}^{d-p} + d \;^{s-1}  \OM_{p-1}^{d-p+1} \rp 
= 0\ .
\]
The triviality of the $Q$-cohomology then implies
\eq
\cas\;^{s-1}  \OM_{p}^{d-p} +  d\;^{s-1}  \OM_{p-1}^{d-p+1}
+ Q\;^{s-2}  \OM_{p}^{d-p+1} = 0 \qquad
(\, p= 0,\dots,d \, ) \ ,
\eqn{bi-desc-0}
where $\;^{s-2}\OM_{p}^{d-p+1}$ is again taken  to be 
a superfield form.
The latter equation is nothing but \equ{bi-descent} with $r=0$. 
The validity of the bi-descent equations for all values of $r$  
is shown by induction, see appendix \ref{proof3-3}. 
\hfill 
\underline{{\em q.e.d.}} 
%%%%%%%%%%%%%%%%%%%%%%%%%%%%%%%%%%%%%%%

The bi-descent equations 
for the pair $(d,s)$ as given by eqs.\equ{bi-descent}  
involve the forms $^{s-r-1}  \OM_{p}^{d-p+r}$ 
which all have the same  total degree
\eq
D=d+s-1 \ .
\eqn{tot-deg}
In the following, we shall consider this number to be fixed to some 
arbitrary value $D \geq 0$. 
For given values of $D, d$ and $s$ related by \equ{tot-deg},
the bi-descent equations for the pair $(d,s)$ 
then become the {\em bi-descent equations for the pair} $(d,D)$,  
i.e. the same set of equations with a different labeling of
ghost- and SUSY-indices:
\eq\ba{c}
\cas\;^{D-p-g}  \OM_{p}^{g} +  d\;^{D-p-g}  \OM_{p-1}^{g+1}
+ Q\;^{D-p-g-1}  \OM_{p}^{g+1} = 0 \es
 (\, p= 0,\dots,d\quad ;\quad g=d-p,\dots,D-p \, )   \ .
\ea\eqn{bi-descent''}
We note that the domain of variation of the indices $p$ and $g$ 
in this set of equations is a
parallelogram $\paral$
in the $(p,g)$ plane
which is bounded by the straight lines 
$p=0$, $p=d$,
$p+g=d$ and $p+g=D$. 
More precisely, each point of $\paral$
represents exactly one of the 
bi-descent equations \equ{bi-descent''},
these equations being parametrized by the form degree and ghost-number of 
the $\cas$-term.
The example $D=3$ is presented in detail in section \ref{app-examples}. 

\medskip 

\noindent
{\bf Summary:}
By definition, the observables of the theory are the 
integrated local functionals 
$\;^s\D_{(d)}$ 
of the form \equ{observ} 
satisfying the cocycle condition \equ{s-cohomol}
and the supersymmetry constraint
\equ{Q-constraints}. 
For a fixed maximal degree 
$D \equiv d+s-1 \geq 0$, they are given by  
superspace integrals of superfield $d$-forms
of ghost-number $0$, i.e. 
\eq
^{D-d+1}  \D_{(d)} = \dint_{M_d}Q\;^{D-d}  \OM_d^0
\qquad 
(\, d = 0 , \dots , D \, ) \ , 
\eqn{obs-d-s}
where $\;^{D-d}  \OM_d^0$ is a non-trivial solution 
of the bi-descent equations 
for the pair  $(d,D)$, i.e. eqs.\equ{bi-descent''}.

%%%%%%%%%%%%%%%%%%%%%%%%%%%%%%%%%%%%%%%%%%%%
%%%%%%%%%%%%%%%%%%%%%%%%%%%%%%%%%%%%%%%%%%%%
%%%%%%%%%%%%%%%%%%%%%%%%%%%%%%%%%%%%%%%%%%%%

%%%%%%%%%%%%%%%%%%%%%%%%%%%%%%%%%%%%%%%%%%%%%%%%%%%
\subsection{Superform solutions of the bi-descent 
equations}\label{superform sol}

When varying $d$ from $0$ to its maximum value $D$, the 
parallelograms $\paral$ 
fill up the triangle $Tri(D)$ of
vertices $(0,0)$, $(D,0)$ and $(0,D)$. 
The points of this triangle describe in a one-to-one fashion 
the {\em bi-descent equations for the forms of total degree $D$},  
i.e. the  bi-descent equations for the pairs $(d,D)$ with 
$d = 0, \dots , D$:
\eq\ba{c}
\cas\;^{D-p-g}  \OM_{p}^{g} +  d\;^{D-p-g}  \OM_{p-1}^{g+1}
+ Q\;^{D-p-g-1}  \OM_{p}^{g+1} = 0 \es
( \, p\ge 0 \quad ,\quad g\ge0\quad ,\quad p+g\le D \, )  \ .
\ea\eqn{bi-descent'}
The bi-descent equations \equ{bi-descent''} 
represent subsets of the latter equations which are 
closed in the sense that each equation of 
\equ{bi-descent'} corresponding to a
point $(p,g)$ $\in$ $\paral$ only involves 
forms corresponding to points of $\paral$, i.e. 
it represents an equation of the set \equ{bi-descent''}. 
Hence a non-trivial solution of \equ{bi-descent'} 
also represents a solution of  \equ{bi-descent''}. 
However, the converse is not necessarily true.
Indeed, two forms $\;^{D-p-g}\OM^g_p$ and
$\;^{D-p-g}(\OM')^g_p$ belonging to the intersection of two
parallelograms $\paral$ and $Par(d',D)$  
might represent different solutions of the two
corresponding sets of bi-descent equations.
%%%%%%%%%%%%%%%%%%%%%%%%%%%%%%%%%%%%

In this subsection, we shall look for solutions of the  
system of equations 
\equ{bi-descent'}, thereby 
providing a special set of solutions
of the bi-descent equations  \equ{bi-descent''}.
The search of the general solution of eqs.\equ{bi-descent''} is
postponed to section \ref{general-solution}.

We first introduce the set of 
$q$-superforms (cf. eq.\equ{s-form})
\eq
\hO_{q}^{D-q} = \dsum{p=0}{q}\;^{q-p}  \OM_p^{D-q}
 (d\theta)^{q-p}  
 \qquad (\, q=0,\dots,D \, )\ ,
\eqn{s-forms}
which contains all of the 
superfield forms appearing in equations \equ{bi-descent'}.
In fact, the superform
$\hO_{q}^{D-q}$ contains the superfield forms of ghost-number $D-q$, 
i.e. those located on the horizontal line $g=D-q$ of the 
triangle $Tri(D)$.
It is easy to check that the equations \equ{bi-descent'} 
are equivalent to the
{\em superdescent equations}
\eq
\cas\hO_{q}^{D-q} + \hd\hO_{q-1}^{D-q+1}=0
\qquad (\, q=0,\dots,D \, ) \ ,
\eqn{s-descent}
where $\hd$ denotes the exterior derivative in superspace, see
eq.\equ{s-ext-der}. 
This equivalence allows us to solve the bi-descent equations 
\equ{bi-descent'} in superspace in terms of
superforms. 
In fact, in this subsection, we shall only be interested in superforms
which are
 polynomials of the basic superforms $\hA(x,\theta)$, $C(x,\theta)$ and
their exterior superderivatives.

The corresponding observables \equ{obs-d-s} are then 
obtained by integrating a non-trivial solution 
$\hO_D^0$ of the superdescent equations \equ{s-descent}
over the collection ${\cal M}$ $=$ $( M_0, M_1, \dots ,$ $M_D)$ of manifolds
according to \equ{s-integral}:
\eq
\hat{\Delta} _{(D)} \equiv \int_{\theta} \int_{{\cal M}} \hO_D^0
= \dsum{p=0}{D} \;^{D-p+1} \D_{(p)} \; (d\theta)^{D-p} 
\ , 
\eqn{dirsum}
where each of the expressions $\;^{D-p+1} \D_{(p)}$  involves another 
component of the superform $\hO_D^0$ (cf.\equ{s-forms}): 
\eq
^{D-p+1} \D_{(p)} = \dint_{M_p} Q \;^{D-p} \OM_p^0
\equiv \dint_{M_p}  \;^{D-p+1} \om_p^0 
 \qquad (\, p = 0, \dots , D \, ) 
  \ .
  \eqn{super-obs}

Here, a solution $\hO$ of  \equ{s-descent}
 is considered to be non-trivial if it 
cannot be written as $\cas\hO'+\hd\hO''$.

Let us now determine the 
non-trivial solutions of the superdescent 
equations \equ{s-descent}, i.e. the elements of the cohomology 
$H(\cas|\hd)$
of the BRST operator $\cas$
modulo the superderivative $\hd$, in the space $\esupsup$
of the local polynomials of
the superconnection $\hA$, the superghost $C$ and 
their $\hd$-derivatives. 
Since the BRST transformations \equ{BRS-super} 
for $\hA(x,\theta)$ and $C(x,\theta)$  
have exactly the same structure as in ordinary pure 
YM theory, the well known results  valid in the latter theory
(see reference~\cite{dubois} and the
reviews~\cite{pig-sor,barnich}) 
can directly be applied after putting ``hats'' on all quantities.
We will use the notation of reference~\cite{barnich}.

First, we introduce the following supercocycles:
\eq\ba{l}
\theta_r(C) = (-1)^{{m_r}-1} \ \dfrac{{m_r}! \, ({m_r}-1)!}{g_r !}
\ \tr C^{g_r} \quad (\, g_r=2m_r-1 \, ) \\[4mm]
f_r(\hF) = \tr \hF^{{m_r}}
\qquad \qquad \qquad \qquad \qquad \quad
\ ( \, r= 1,\dots, \mbox{rank }G  \, )\ .
\ea\eqn{s-cohom}
Here, $\hF=\hd\hA+\hA^2$ is the curvature
of the superconnection $\hA$ and the
index $r$ labels the $r^{\rm th}$ Casimir
operator of the structure group (gauge group) $G$, whose degree is
denoted by $m_r$.
The cocycles (\ref{s-cohom}) 
are related by superdescent equations involving 
superforms $\lc \hat\theta_r \rc_p^{g_r-p}$
of form degree $p\geq 0$ and ghost-number $g_r-p$:
\eq\ba{c}
\cas \lc \hat\theta_r \rc_p^{g_r-p} 
+ \hd \lc \hat\theta_r \rc_{p-1}^{g_r-p+1}
=0
\qquad (\, p=0,\dots, g_r 
\, ) 
\ ,\es
\mbox{with}\quad 
\lc \hat\theta_r \rc_0^{g_r} = \theta_r(C)
\quad {\rm and} \quad 
\hd \lc \hat\theta_r \rc^0_{g_r} = f_r(\hF)\ .
\ea\eqn{desc-theta}
According to the last equation, the ``top'' superform 
$\lc \hat\theta_r \rc^0_{g_r}$ is the 
Chern-Simons superform of degree $g_r$ 
associated to the $r^{\rm th}$
Casimir operator.
 
Obviously, \equ{desc-theta} corresponds 
to the superdescent
equations \equ{s-descent}
and thus yields a solution of the latter equations. 
More general solutions are found 
by multiplying the cocycle $\theta_r (C)$
by a certain number of factors $f_s(\hF)$ since
the latter are both $\cas$- and $\hd$-invariant. 
Thus, we introduce 
the following supercocycle (belonging to the $\cas$-cohomology  in the
space $\esupsup$):
\eq\ba{c}
\hHH_R \equiv \theta_{r_1} (C) 
f_{r_2}(\hF)\cdots f_{r_L}(\hF)\ ,\es
\mbox{with} \quad 
L\geq 1 \ , \quad r_i \le r_{i+1}\ .
\ea\eqn{s-d-cohom}
This cocycle is 
of ghost-number $g_{r_1}$ and superform 
degree $D_R= \dsum{i=2}{L} \, 2m_{r_i}$ (cf.\equ{s-cohom}).
By virtue of equations \equ{desc-theta}, the superforms 
\eq
\hO^{g_{r_1}-p}_{D_R+p} = \lc{\hat\theta}_{r_1} \rc^{g_{r_1}-p}_p
f_{r_2}(\hF)\cdots f_{r_L}(\hF)
\quad (\,
p=0,\dots,g_{r_1} \; ; \ L\ge1 \,)  
\eqn{3.x}
obey the superdescent equations
\eq
\cas\hO_{D_R+p}^{g_{r_1}-p} + \hd\hO_{D_R+p-1}^{g_{r_1}-p+1} = 0
\qquad (\, p  = 0,\dots, g_{r_1} \, ) \ .
\eqn{s-descent-part}
The most general solution of the superdescent equations \equ{s-descent}
in the space $\esupsup$ is obtained by considering a supercocycle
of the form (\ref{s-d-cohom}) which is non-linear in the monomials
$\theta _r (C)$. However, in view of the construction of observables 
(which have zero ghost-number by definition), 
we are only interested in the most general solution containing 
superforms of ghost-number $0$ and the latter are given by 
(\ref{3.x}) according to the results of section 10.7
of reference~\cite{barnich}, adapted to the present superspace
formalism.

The corresponding observables are now constructed according to 
(\ref{dirsum})(\ref{super-obs}), by using \equ{3.x} for $p=g_{r_1}$, i.e. 
\eq\ba{rl}
& \hO^{0}_{D} = \lc{\hat\theta}_{r_1} \rc^{0}_{g_{r_1}}
f_{r_2}(\hF)\cdots f_{r_L}(\hF)
 \equiv \dsum{p=0}{D} \;^{D-p} \Omega ^0_p \,(d\theta)^{D-p} 
\ , \es
\mbox{with} & D=D_R+g_{r_1} 
= 2 \, \dsum{i=1}{L} \, m_{r_i} -1 
\ , \quad L \geq 1 \ .
\ea\eqn{3.xx}
Note that $D$ is necessarily odd. 

%%%%%%%%%%%% %%%%%%%%%%%%%%%%%%%%%%%%%
There is an alternative way of writing the observables
which amounts to a simpler manner of 
extracting the polynomials $^{D-p+1} \om _p^0 (x)$
from the superform $\hO ^0_D$. This procedure is 
suggested by the fact that the
exterior derivative $\hd = d\theta \, \partial_{\theta} +d$
differs from the superspace SUSY operator $\partial _{\theta}$ 
by a factor $d\theta$ and the addition of a spacetime derivative. 
Indeed, let us consider the exterior derivative of $\hO ^0_D$ and
write its expansion with respect to $d\theta$
(see (\ref{s-curvature})(\ref{expsi})):
\eq\ba{c}
\hd\hO ^0_D =
f_{r_1}(\hF)  \cdots  f_{r_L}(\hF) 
= f_{r_1}(F_A) \cdots f_{r_L}(F_A) 
+ \dsum{p=0}{D}  \;^{D+1-p}W^0_{p}  \; (d\theta)^{D+1-p}  \ , 
\ea\eqn{exp-hd-OM}
with 
\begin{eqnarray*}
\;^{D-p+1} W ^0 _p 
\!\!\! &=& \!\!\! 
Q \;^{D-p} \Omega ^0 _p + d \;^{D-p+1} \Omega ^0 _{p-1}
\\
\!\!\! & = & \!\!\! 
\; ^{D-p+1} \omega ^0 _p (x) 
+ d \; ^{D-p+1} \Omega ^0 _{p-1} (x, \theta ) \ .
\end{eqnarray*}
Henceforth, in the integral \equ{super-obs}
which yields the observables, 
we can substitute the form $\;^{D-p+1}\om _p$
by
 the value of $\;^{D-p+1}W _p$ at $\theta =0$:
\eq\ba{c}
^{D-p+1}\D_{(p)} = \dint_{M_p}  \;^{D-p+1}w_p^0 
\qquad \qquad (\, p = 0, \dots , D \, )\ ,\es
\mbox{with}\quad  \;^{D-p+1}w_p^0  = 
\left. \;^{D-p+1}W_p^0\right|_{\theta=0}   \ .
\ea\eqn{super-obs'}
 
Before concluding, 
we note that application of the superderivative
$\hd$ to \equ{exp-hd-OM} and use of its nilpotency, leads to 
\[
Q \;^{D-p+1} W^0_p + d\;^{D-p+2} W^0_{p-1} = 0 
 \qquad (\, p = 0, \dots , D \, )
\ ,
\]
which, taken at $\theta =0$, yields 
\eq
Q \;^{D-p+1} w ^0_p + d\;^{D-p+2} w ^0_{p-1} =0 
\qquad (\, p = 0, \dots , D \, )
\ .
\eqn{Witten-Q}
These relations for the integrands $\;^{D+1-p} w^0 _{p}$
of the observables \equ{super-obs'} 
are nothing but Witten's descent equations 
in a general supergauge (generalizing 
eqs.\equ{wdes} which hold in the WZ-gauge and  
involve $\tQ$ rather than $Q$). 
In the WZ-gauge, the polynomials $\;^{D+1-p} w^0 _{p}$ reduce 
- by construction - to the Donaldson-Witten polynomials discussed in
subsection 2.4.2. 
In particular, in the WZ-gauge, we obtain 
\eq
^{D+1}\om^0 _0 
= f_{r_1}(\f) \cdots f_{r_L}(\f)
\qquad {\rm with} \ \  
L \geq 1 , \quad  r_i\le r_{i+1}
\ ,
\eqn{Witten-0-form}
i.e.  Witten's well-known result~\cite{witten} that 
the algebra of observables is generated, at form degree zero, by the
invariant monomials $f_s(\f)$. 
The examples $D=1$ and $D=3$ will be presented in more detail 
in section \ref{app-examples}. 

Anticipating the discussion of the next subsection, which
shows that there are no other {\em non}-trivial observables, 
we can summarize our results as follows.

\noindent{\bf Summary:} 
Apart from the `trivial' observables (\ref{trivobserv}), 
there exist further ones. 
All of these observables, as defined by the conditions 
\equ{s-cohomol} and \equ{Q-constraints}, are
given by eqs.\equ{super-obs'}.
In the latter expressions, 
the superfield forms $\;^{D-p+1}W_p^0$ are 
the coefficients appearing in the expansion 
\equ{exp-hd-OM}, the superform $\hO_D^0$ 
being the non-trivial solution \equ{3.xx} of the 
superdescent equations \equ{s-descent} in the space $\esupsup$ of 
polynomials
in the basic superforms $\hA$, $C$ and their exterior superderivatives.
The forms $^{D-p+1}w^0_p$ satisfy the generalization
of Witten's descent equations to a general supergauge, 
i.e. eqs.\equ{Witten-Q}.

%%%%%%%%%%%%%%%%%%%%%%%%%%%%%%%%%%%%%%%%%%%%%%%%%%%%%
\subsection{General solution of the bi-descent 
equations for the pair $(d,D)$}\label{general-solution}

As noted at the beginning of the last section,  
the solutions of the superdescent equations \equ{s-descent} in the space
$\esupsup$ (which is generated by the 
superforms $\hA , \, C$ and the operators 
$\cas, \, \hd$) represent  
{\em a priori} only a special set of solutions of the 
bi-descent equations for the pair $(d,D)$, 
i.e. eqs.\equ{bi-descent''}. 
Henceforth, we have to determine the general
non-trivial solution of the latter equations in order to obtain
the general set of observables. 
At this point, we only state and comment on the main result,  
leaving the proof  for appendix \ref{app-gen-sol}.
%%%%%%%%%%%%%%%%%%%%%%%%%%%%%%%%%%%%%%%%%%%%
\begin{enumerate}\item[]
{\bf Proposition \ref{observables in SF form}.4}
The general solution of the bi-descent equations \equ{bi-descent''}
for the pair $(d,D)$ is generated, at ghost-number zero,  by two
classes of solutions. The first one is given by the superfield
forms 
\eq\ba{rl}
\;^{D-d}\OM^0_d 
\, (d\theta )^{D-d} 
\!\!\! & =\lc \lc\hat\theta_{r_1}\rc^0_{g_{r_1}} 
f_{r_2}(\hF) \cdots f_{r_L}(\hF) \rc_{s=D-d,\,p=d} \es
\qquad \qquad \qquad \qquad 
\mbox{with}  \ \  D\!\!\! &= 2\, \dsum{i=1}{L} m_{r_i} -1 
\ , \quad L\ge1\ ,
\ea\eqn{gen-sol-1}
where the Chern-Simons superform $\lc\hat\theta_{r}\rc^0_{g_{r}}$
and supercurvature invariant $f_{r}(\hF)$ are defined
by (\ref{desc-theta})(\ref{s-cohom}).

The second class of solutions depends on the 
superfield forms $F_A, \bpsi$
and $\bfi$ defined in eqs.\equ{s-curvature}\equ{s-field-red}
and it is given by 
\eq
\;^{D-d}\OM^0_d = 
\;^{D-d}\ZZ^0_d(F_A,\bpsi,\bfi,{D_A}\bpsi,{D_A}\bfi) \ .
\eqn{gen-sol-2}
Here, 
$\;^{D-d}\ZZ^0_d$ is an arbitrary invariant polynomial of its
arguments, 
which has a form degree $d$ and SUSY-number $D-d$
and which is non-trivial in the sense that 
$\;^{D-d}\ZZ ^0_d \not= d \;^{D-d} \Phi_{d-1}^0 + Q  \;^{D-d-1}
\Phi_{d}^0$. 
\end{enumerate}
%%%%%%%%%%%%%%%%%%%%%%%%%%%%%%%%%%
\noindent{\bf Proof:} See appendix \ref{app-gen-sol}.

Concerning the invariant forms \equ{gen-sol-2}, 
we note that they represent
the general cohomology classes of the BRST operator $\cas$ in
the space $\esup$ by virtue of a mere 
adaptation of the results of section 8
of reference~\cite{barnich}.

According to (\ref{super-obs}), 
the observables are obtained as integrals
of the $Q$-variation of the 
solutions \equ{gen-sol-1}\equ{gen-sol-2} 
over a $d$-dimensional  manifold.
The ones corresponding to \equ{gen-sol-1} 
coincide with the corresponding expressions 
calculated in section \ref{superform sol} 
(i.e. \equ{super-obs'} with $p=d$) since the 
corresponding integrands only differ by a total derivative.
In fact, for $d=0,1,\dots,D$, the superfield forms $\;^{D-d}\OM^0_d$
given in \equ{gen-sol-1} are nothing but those introduced in \equ{3.xx}.
Hence, the solutions \equ{gen-sol-1} provide the same observables
as the superform solutions \equ{3.xx}.

On the other hand, 
for the solutions \equ{gen-sol-2}, one gets the integrals 
\eq
\;^{D-d+1}\D_{(d)} = 
\int_{M_d} Q \;^{D-d}\ZZ^0_d(F_A,\bpsi,\bfi,{D_A}\bpsi,{D_A}\bfi)\ .
\eqn{triv-obs}
As pointed out after equation \equ{susy-redef},  
the operator $Q$ simply reduces to $Q_0$ when acting 
on an invariant polynomial $\;^{D-d}\ZZ^0_d$. 
Yet, as noted after equation \equ{nilp}, the action of  $Q_0$
is isomorphic to the one of the operator $\tQ$ 
describing supersymmetry transformations in the WZ-gauge.
This means that the solution \equ{triv-obs}, 
if written out in the WZ-gauge, is simply  
the $\tQ$-variation of a gauge invariant polynomial.
Thus, it is 
{\it trivial} in the sense of equivariant cohomology , see
eqs.(\ref{Q-cohom})-(\ref{qnontriv}).
This explicitly shows that the cohomology  defined by equations 
\equ{s-cohomol}-\equ{Q-constraints} is {\it not} 
equivalent to the equivariant cohomology: the
difference is precisely given by the expressions of the form
\equ{triv-obs}, which are manifestly $\tQ$-trivial.

We conclude that, apart from the solutions \equ{gen-sol-2} which
are ``equivariantly'' trivial, the {\em general} solution of the bi-descent
equations that we described in this section 
does not yield any more solutions than 
those obtained in terms of {\em superforms} in section~\ref{superform sol}.
In other words, the solution constructed by 
using superforms represents 
the most general, equivariantly non-trivial expression for the observables.

%%%%%%%%%%%%%%%%%%%%%%%%%%%%%%%%%%%%%%%%%%%%%%%%%%%%%%5
\section{Explicit expressions}\label{app-examples}

%%%%%%%%%%%%%%%%%%%%%%%%%%%%%%%%%%%%%%%%%%%%%%%%%%%%%%%
\subsection{An example of bi-descent and superdescent\\
 equations}\label{app-bi-desc}

There is a graphical way of representing the 
sets of bi-descent equations 
which allows us to exhibit explicitly the combinatorics 
leading to the superdescent equations \equ{s-descent}. 

By way of illustration, let us 
 consider the case of total degree $D=3$.
We have 10 superfield forms $^{3-p-g} \OM_p^g$ 
in the bi-descent equations 
for total degree $D=3$, i.e. eqs.\equ{bi-descent'},  
which can be represented in the 
$(p,g)$ diagram: 
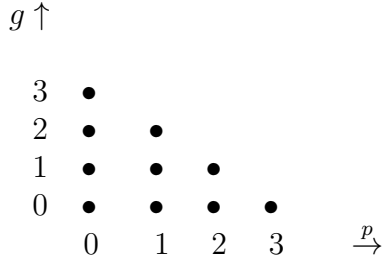
\[
\matrix{
\!\!\!\!g \uparrow &&&&&\cr
         &&&&&\cr
        3&\eqbul\ &       &     &     &\cr
        2&\eqbul\ &\eqbul  &     &     &\cr
        1&\eqbul\ &\eqbul  &\eqbul&     &\cr
        0&\eqbul\ &\eqbul  &\eqbul&\eqbul& \cr
         &\!\!\!0&1&2&3&  \quad \stackrel{p}{\rightarrow} }
\]
E.g. the point on the outer right represents the form 
$^0 \OM_3^0$. 
The forms listed in the previous diagram appear in different 
sets of  bi-descent equations 
\equ{bi-descent''}: 
for $d= 0, 1, 2 ,3$ and $D=3$,
the latter bi-descent equations correspond to the following 
 sub-diagrams of the previous diagram: 
%%%%%%%%%%%%%%%%%%%%%%%%%%%%%%%%%%%%%%%%%%%%%%%%%%%%%
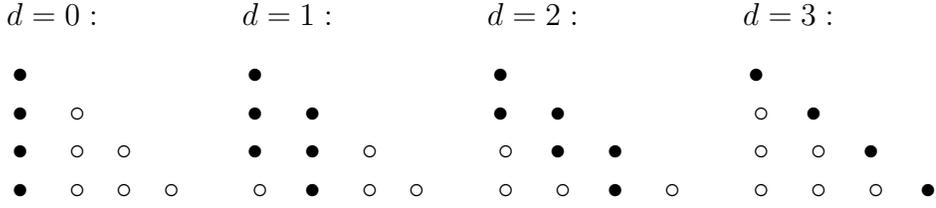
\[\ba{llll}
d=0: \qquad
& 
d=1: \qquad
& 
d=2: \qquad
& 
d=3:
\es
\matrix{
\eqbul &       &      &      & \cr 
\eqbul &\circ  &      &      & \cr
\eqbul &\circ  &\circ &      & \cr
\eqbul &\circ  &\circ &\circ & }
&
\matrix{
\eqbul &       &      &      & \cr 
\eqbul &\eqbul &      &      & \cr
\eqbul &\eqbul &\circ &      & \cr
\circ  &\eqbul &\circ &\circ & }
&
\matrix{
\eqbul &       &      &      & \cr 
\eqbul &\eqbul &      &      & \cr
\circ  &\eqbul &\eqbul &      & \cr
\circ  &\circ  &\eqbul &\circ & }
&
\matrix{
\eqbul &       &      &      & \cr 
\circ &\eqbul &      &      & \cr
\circ  &\circ &\eqbul &      & \cr
\circ  &\circ  &\circ &\eqbul & }
\ea\]
%%%%%%%%%%%%%%%%%%%%%%%%%%%%%%%%%%%%%%%%%%%%%%%%%%%%%%%%%%%%
Thus, one clearly sees how the parallelograms representing 
equations 
\equ{bi-descent''} 
overlap for the various values of $d$ (and a fixed value $D$) 
%and $s$ satisfying $d+s-1 =D$, 
finally covering the full triangle in the $(p,g)$ plane
which represents the bi-descent equations \equ{bi-descent'}.
Obviously, this triangle also represents the 
superdescent equations \equ{s-descent}.
The latter equations presently 
take the same form as the descent equations in   
$3$-dimensional Chern-Simons field theory for which the solution is 
well known, e.g. see reference \cite{pig-sor}.  Thus, 
for  $D=3$, the non-trivial
 solution 
 of the superdescent equations \equ{s-descent} is given by
\eq\ba{l}
\hO_3^0 =  \tr ( \hA \hd\hA + \dfrac{2}{3} \hA^3 ) \ , \quad
\hO_2^1 =  \tr (\hA \hd C ) \ , \quad
\hO_1^2 =  \tr ( C \hd C ) \ , \quad
\hO_0^3 =  -\frac{1}{3} \; \tr C^3 \ .
\ea\eqn{super-D=3}

%%%%%%%%%%%%%%%%%%%%%%%%%%%%%%%%%%%%%%%%%%%%%%%
\subsection{Some examples of observables}\label{app-b}

In this section, we consider the structure group $U(1) \times SU(2)$ 
to illustrate the conclusions of section \ref{observables in SF form}. 
For this group, there are two Casimir operators: the $U(1)$ generator 
itself (the charge) and the
quadratic Casimir of $SU(2)$. Their degrees are respectively $m_1=1$ and
$m_2=2$. 

In the sequel, we shall use an index `(a)' for `abelian'.
The ghosts, connections and curvatures are, respectively,   
given by 
the following superfields and -forms: 
\[\ba{llll}
U(1): \quad&C_{\rm(a)}\ ,\quad&\hA_{\rm(a)}\ ,
\quad& \hF_{\rm(a)}= \hd\hA_{\rm(a)} \es
SU(2): \quad&C\ ,\quad&\hA\ , \quad& \hF= \hd \hA+\hA^2\ .
\ea\]
The ``canonical'' basis \equ{s-cohom}
of the cohomology $H(\cas)$ reads as 
\eq\ba{ll}
\theta_1 =  \theta_1(C) = C_{\rm(a)}\ ,
\quad &f_1 = f_1 (\hF_{\rm(a)} ) 
= \hF_{\rm(a)} \es
\theta_2  = \theta_2(C) = -\frac{1}{3}\tr C^3\ ,
\quad &f_2 = f_2(\hF) = \tr \hF^2 
\ea\eqn{a-s-coho}
and the canonical descent equations \equ{desc-theta} involve the forms
\eq\ba{lll}
U(1): \quad &[\hat\theta_1]_{0}^{1}=\theta_1\ , 
& \quad [\hat\theta_1]_{1}^{0}= \hA_{\rm(a)} \es
SU(2): \quad&
[\hat\theta_2]_{0}^{3}= \theta_2\ ,
& \quad [\hat\theta_2]_{1}^{2} = \tr(C\hd C) \es
\quad&[\hat\theta_2]_{2}^{1}=\tr(\hA\hd C)\ ,
& \quad [\hat\theta_2]_{3}^{0}= \tr(\hA\hd\hA+\frac{2}{3}\hA^3)\ .
\ea\eqn{a-desc}
Let us now look for the observables that can be 
deduced from the cohomology of $\cas$
modulo
 $\hd$, i.e. from the non-trivial solutions of
the superdescent equations \equ{s-descent}. We shall consider
three cases, namely the two sets of basic observables corresponding to the 
two Casimir operators, and one set of ``composite'' observables. 

An example of a solution of the superdescent equations
which does {\em not} yield observables is obtained from the  
bottom superform $\hO_0^4 = \theta_1\theta_2$: this provides 
a simple illustration of the general formalism
where the climbing stops at the form degree equal to the
ghost-number of $\theta_1$, namely degree $1$.

%%%%%%%%%%%%%%%%%%%%%%%%%%%%%%%%%%
\subsubsection{Solution corresponding to the Casimir of 
$U(1)$}\label{Casimir U(1)}

The superdescent equations for total degree $D=1$, 
\[ 
\quad \cas\hO_0^1 = 0 \quad , \quad 
 \cas\hO_1^0 + \hd\hO_0^1 = 0  \ ,
\]
are solved by the superforms 
\[
\ba{ll}
\hO_0^1 \, = & ^0  \OM_0^1 = \theta_1 = C_{\rm(a)}
\es
\hO_1^0 \, = & ^0  \OM_1^0 \, + \; ^1  \OM_0^0 \, d\theta 
\, = \,   \hA_{\rm(a)} \ . 
\ea
\]
The latter coincide with the ``canonical'' 
superforms \equ{a-desc} for $U(1)$.

According to 
the results of section 3.3 (see eqs.(\ref{exp-hd-OM})(\ref{super-obs'})),
the observables are obtained from the superspace exterior derivative
of the top superform $\hO_1^0$ (which has ghost-number zero)
and given by the expansion at $\theta=0$:
\eq\ba{l}
\left. \hd \hO_1^0 \right| _{\theta=0} 
=  \left. \hF_{\rm(a)} \right| _{\theta=0} 
= F_{\rm(a)}
+ \;^1w^0 _1 \, d\theta + \;^2w^0_0 \, (d\theta)^2
\ , \es 
\mbox{with} 
 \ \ \;  F_{\rm(a)} = da_{\rm(a)} \ , 
\ \; \;^1w^0 _1 =   \p_{\rm(a)} + d\chi_{\rm(a)}
\ , \ \; \;^2w^0 _0 = \f_{\rm(a)}.
\ea\eqn{abel-observ}
 Apart from the `trivial' observable $\int_{M_2} F_{\rm(a)}$, 
the observables are the integrals of the forms 
$\;^1w^0 _1$, 
$\;^2w^0 _0$ on closed submanifolds $M_1$ and $M_0$, respectively.  
The polynomials $w_p$ satisfy Witten's descent equations 
in a general supergauge, i.e. eqs.\equ{Witten-Q}. 

Equivalently -- cf.(\ref{dirsum})-(\ref{super-obs}) --
the superspace integral of the superform $ \hO_1^0 $ 
over a collection $\MM = (M_0, M_1)$ 
of closed submanifolds
is a direct sum of two integrals,  
\begin{equation}
\label{multi-int-1}
\hat{\Delta} _{(1)} \equiv 
\int_{\theta} \dint_{\MM} 
 \hO _1^0
= 
\dint_{\MM} 
\dth \hO _1^0
= \dsum{p=0}{1} (d\theta)^{1-p} 
  \dint_{M_p}\;^{2-p}  \om^0 _p \ ,
\end{equation}
where the $p$-forms $\;^{2-p}  \om^0 _p$ are the coefficients of the
expansion of $\dth\hO^0_1$.
Each integral in (\ref{multi-int-1}) 
defines an expression belonging to the 
SUSY-constrained cohomology of $\cas$. These integrals 
coincide with 
those of the forms $w_p$ defined in \equ{abel-observ}.

In the WZ-gauge
$\chi=0$,
the expressions \equ{abel-observ} reduce to the 
Donaldson-Witten polynomials generated from the 
invariant $\f_{\rm(a)}$ using the supersymmetry operator
$\tilde Q$, see eqs.(\ref{expsi}) with $m=1$. 
In our approach, these polynomials have been 
generated for $D=1$ from the bottom 
superform $\hO_0^1 = C_{\rm(a)}$ which solves superdescent equations.

%%%%%%%%%%%%%%%%%%%%%%%%%%%%%%%%%%
\subsubsection{Solution corresponding to the Casimir of 
$SU(2)$}\label{Casimir SU(2)}

The bottom form $\hO_0^3 = \theta_2$ has total degree $D=3$
and the solution of the superdescent equations
is given by 
the superforms \equ{super-D=3}. These expressions coincide with the 
canonical 
superforms \equ{a-desc} for $SU(2)$.
Applying again proposition 4, we obtain the observables from the
expansion
\eq
\left.\hd \hO_3^0 \right|_{\theta=0} 
= 
\left. \tr \hF ^2  \right|_{\theta=0} 
= \tr {F_a} ^2 + 
\dsum{p=0}{3} \;\;^{4-p}  w^0_p \; 
 (d\theta)^{4-p} 
\ ,
\eqn{hat-om-3-0}
the last term being an exterior derivative.
By substituting the component field expansions \equ{agf}\equ{aac} 
of $\hA$ into $\hO_3^0$, we obtain the following 
explicit expressions
for the spacetime forms: 
\eq\ba{l}
^4w^0 _0 = \tr(\f^2 + 2\f\chi^2 ) \es
^3w^0 _1 =  \tr 2 ( \psi\f + \p\chi^2 + 
\f D_a\chi ) 
+ d \, \tr ( \frac{2}{3} \chi^3 )  \es
^2w^0 _2 = \tr(\p^2 + 2\f F_a + 2\p D_a \chi) 
+  d \, \tr (\chi D_a \chi ) \es
^1w^0 _3 = \tr( 2 \p F_a ) 
+d \, \tr (2 \chi F_a ) ) \ .
\ea\eqn{list-fi2}
 The observables are the integrals of these forms 
 (and of $\tr {F_a} ^2$) 
 on closed submanifolds of appropriate dimension. 

In the WZ-gauge $\chi=0$,
the expressions for the observables again 
reduce to Witten's result (generated from the quadratic 
invariant $\tr\f^2$), i.e.  
eqs.(\ref{expsi}) with $m=2$.

%%%%%%%%%%%%%%%%%%%%%%%%%%%%%%%%%%%%%%%%%%%
\subsubsection{An example of ``composite observables''}\label{compos obs}

As stated at the end of
section \ref{superform sol}, all other
observables are integrals 
whose integrands are polynomials of the forms $w_p$ 
that we constructed in the last two subsections
(i.e. of the forms associated to the Casimir operators). 
Let us illustrate this with the simplest example,
generated by the bottom form $\hO^1_2=\theta_1 f_1$ which is of
total degree $3$. The corresponding
top superform is given by 
$\hO_3^0= \hA_{\rm(a)} \hF_{\rm(a)}$ and the expansion of its 
superderivative
$\hd\hO_3^0$ = $(\hF_{\rm(a)})^2$
yields the following integrands
for the observables: 
\eq\ba{l}
^4 {\tilde w} _0^0 = ( ^2w_0^0 )^2
 =  \f_{\rm(a)} ^2  
\es
^3 {\tilde w} _1^0 = 2 \, (^1w_1^0)\, ( ^2 w_0^0 )  \es
^2 {\tilde w}_2^0 =  2 \; ^2w_0^0 \; F_{\rm(a)} + (^1w_1^0)^2 \es
^1 {\tilde w}_3^0 = 2 \; ^1w_1^0 \; F_{\rm(a)} \es
^0 {\tilde w}_4^0 = F_{\rm(a)}^2 \ .
\ea\eqn{obsII}
Obviously, these forms are polynomials in the
basic forms given in eqs.\equ{abel-observ}
and in the abelian curvature invariant $f_1$ = $F_{\rm(a)}$.

%%%%%%%%%%%%%%%%%%%%%%%%%%%%%%%%%%%%%%%%%%%%%%%
\section{Concluding remarks}

We have shown that the problem of determining the equivariant
cohomology of topological Yang-Mills theories can be reduced to that of
computing the Yang-Mills BRST cohomology (modulo $d$) in the space of
polynomials depending on the components of the
Yang-Mills superconnection $\hA$, its superghost $C$ and their
exterior derivatives -- all these components being superfields. 
The determination of this cohomology relies on different extensions 
of well-known techniques~\cite{barnich}, on one hand to superspace, 
and on the other hand to 
the case where one has two BRST-like operators, namely $\cas$ and $Q$. 
This leads to the consideration of ``bi-descent equations''
generalizing the usual descent equations.

Our main result is the following one. Apart from solutions of the  
bi-descent equations
that are trivial in the sense of equivariant cohomology   
(i.e. the trivial observables determined by 
\equ{gen-sol-2}), the general
non-trivial solution \equ{gen-sol-1} of these equations 
(describing an observable 
$\;^s \Delta _{(d)}$ of dimension $d$ and SUSY-number $s$) 
is given as the superspace integral 
$\int_{M_d} d\theta \, \;^{s-1} \Omega_d^0$, where 
$\;^{s-1} \Omega_d^0$ is 
a coefficient of some superform which has total degree $D=d+s-1$, 
this superform being a solution of
a set of ``super-descent equations''. 
In other words, the observables
are determined by the cohomology of the BRST operator (modulo the
exterior superderivative $\hd$) in the space of 
superforms which are polynomials in the
superconnection $\hA$, the Faddeev-Popov 
ghost-superfield $C$ and their exterior superderivatives.
When specialized to the Wess-Zumino gauge, our result 
reproduces Witten's observables~\cite{witten}. 
The generalization of our approach to more complex models 
is currently under study and 
will be reported upon elsewhere~\cite{clisthen,boldo2}.

%%%%%%%%%%%%%%%%%%%%%%%%%%%%%%%%%%%%%%%%%%%%%%

\bigskip 

\noindent{\bf Acknowledgments:} 
O.P. thanks the Institut de Physique Nucl\'eaire of the Universit\'e 
Claude Bernard, Lyon, and all its staff for its very kind hospitality
during a stay  which has been at the origin of this work.
F.G. acknowledges discussions with Fran\c cois Delduc. C.P.C thanks 
LPTHE (Universit\'e Paris 6) and the Abdus Salam Institute -- ICTP 
(Trieste) for hospitality and as well ICTP for an Associate Fellowship. 

%%%%%%%%%%%%%%%%%%%%%%%%%%%%%%%%%%%%%%%%%%%%%%%%%
%%%%%%%%%%%%%%%%%%%%%%%%%%%%%%%%%%%%%%%%%%%%%%%%%
%%%%%%%%%%%%%%%%%%%%%%%%%%%%%%%%%%%%%%%%%%%%%%%%%
%%%%%%%%%%%%%%%%%%%%%%%%%%%%%%%%%%%%%%%%%%%%%%%%%
\appendix

%%%%%%%%%%%%%%%%%%%%%%%%%%%%%%%%%%%%%%%%%%%%%%%%%
%%%%%%%%%%%%%%%%%%%%%%%%%%%%%%%%%%%%%%%%%%%%%%%%%

\section{Proofs of some propositions and lemmas}\label{app}

%%%%%%%%%%%%%%%%%%%%%%%%%%%%%%%%%%%%%%%%%%%%%%%%%
\subsection{Proof of Proposition 
\ref{observables in SF form}.1}\label{proof3-1} 
%%%%%%%%%%%%%%%%%%%%%%%%%%%%%%%%%%%%
%%%%%%%%%%%%%%%%%%%%%%%%%%%%%%%%%%%%

The proof of the results (\ref{lemma1b})(\ref{lemma1b'}) and
(\ref{L1b})(\ref{L1b'})
is based on the triviality of the cohomologies $H(Q)$ and
$H(d)$ for the functional spaces 
$\eord$ and  $\esup$, respectively, see propositions 
\ref{symetries}.1 and \ref{symetries}.2.
Here, we outline 
the proof of (\ref{lemma1b})(\ref{lemma1b'}),
the one of  (\ref{L1b})(\ref{L1b'}) being analogous.

Equation \equ{lemma1b}  
follows from the cocycle 
condition \equ{lemma1a} by virtue of a corollary of 
theorem 9.2 of ref.~\cite{barnich}.  
In the present context, this corollary states that, if the
cohomologies $H(Q)$ and $H(d)$ are both trivial, and if a form
$\;^{s}\om_{p}$ is  $Q$-invariant modulo $d$ (i.e. condition
\equ{lemma1a} holds), then  $\;^{s}\om_{p}$ is $Q$-exact modulo $d$,
i.e. \equ{lemma1b} holds. The same corollary, with the roles of $Q$
and $d$ interchanged, implies that $\;^{s+1}\om_{p-1}$ is $d$-exact
modulo $Q$, i.e.  
\eq
^{s+1}  \om_{p-1} = d\;^{s+1}  \chi_{p-2}  +  Q\;^{s}  \chi_{p-1}  \ .
\eqn{lemma1c}
By substituting the expressions \equ{lemma1b} and  \equ{lemma1c}
into \equ{lemma1a},
we obtain the equation
\[
dQ\lp \;^s\chi_{p-1} -  \;^s\vf_{p-1} \rp =0\ ,
\]
which, due to the triviality of the cohomology $H(d)$, implies 
\[
Q\lp \;^s\chi_{p-1} -  \;^s\vf_{p-1} \rp   + d\;^{s+1}\chi_{p-2} 
= 0\ .
\]
 This $Q$ modulo $d$ invariance condition is solved by  
\[
\;^s\chi_{p-1} = \;^s\vf_{p-1} + Q \;^{s-1}\xi_{p-1} + d \;^{s}\xi_{p-
2}\ .
\]
Introducing this result into eq.\equ{lemma1c} and defining
\[
\;^{s+1}\vf_{p-2} = \;^{s+1}\chi_{p-2} -Q\;^{s}\xi_{p-2}
\]
finally yields the result \equ{lemma1b'}.

%%%%%%%%%%%%%%%%%%%%%%%%%%%%%%%%%%%
%%%%%%%%%%%%%%%%%%%%%%%%%%%%%%%%%%

%%%%%%%%%%%%%%%%%%%%%%%%%%%%%%%%%%%%%%%%%%%%%%%%%%%%%%%%%%%%%
\subsection{Proof of Lemma 1}\label{proof3-2} 

The proof proceeds by induction.
Equation \equ{th-deriv} already holds at form degree $d$.
Let us assume relation \equ{th-deriv}
to be true at form degree $p+1$
and show its validity at degree $p$.  
By applying $Q$ to the 
descent equation \equ{desc-s-d} at degree $p+1$ and using the 
induction hypothesis, we find 
\[
d Q  \;^s  \om_{p}^{d-p} = 0\ .
\]
Due to the triviality of the $d$-cohomology, 
this equation implies 
the $Q$ modulo $d$ cocycle condition
\[
 Q  \;^{s}  \om_{p}^{d-p} + d \;^{s+1}  \om_{p-1}^{d-p} = 0 \ .
\]
According to Proposition \ref{observables in SF form}.1, 
the general solution of the latter is
\[
^s  \om_p^{d-p} = 
 Q  \;^{s-1}  \om_{p}^{d-p} + d \;^{s}  \om_{p-1}^{d-p}\ .
\]
Discarding the derivative term since it does not contribute to the 
descent
equation \equ{desc-s-d} at degree $p+1$, 
we thus obtain the 
result \equ{th-deriv} after replacing 
the spacetime form
$\;^{s-1}  \om_{p}^{d-p}$ by a   
superfield form 
$^{s-1}  \OM_p^{d-p}$ 
thanks to the 
argument leading from \equ{Q-omega}
to \equ{Q-superfield}.
 
%%%%%%%%%%%%%%%%%%%%%%%%%%%%%%%%%%%%%%%%%%%%%%%%%
\subsection{Proof of Proposition 
\ref{observables in SF form}.3}\label{proof3-3} 
   
The first part of the proof was already 
presented after Proposition \ref{observables in SF form}.3.

In order to 
prove the validity of the bi-descent
equations \equ{bi-descent} for all values of $r$
by induction, 
it is convenient to
use the formalism of ``extended forms''~\cite{ext-forms}. 
The latter involve superfield forms of the same total degree,
but of different form degrees and ghost-numbers.
In general, an extended form $\tO_{d+r}$ is supposed to be of the 
form $\tO_{d+r} = \Omega_{d+r}^0 + \Omega_{d+r-1}^1 + \cdots + 
\Omega_0^{d+r}$, but, for the present application, we truncate
the expansion so as to have $d$ as highest form degree, i.e. we
consider
\eq
\;^{s-1-r}\tO_{d+r} = \dsum{p=0}{d} \;^{s-1-r}\OM_{p}^{d-p+r}\quad
(\, r=0,\dots , s -1 \, )\ .
\eqn{ext-forms}
The ``extended differential'' acting on these extended forms is
defined by
\eq
\td=\cas+d
\eqn{ext-d} 
and it is nilpotent $(\td\,{}^2=0)$.

The set of bi-descent equations \equ{bi-descent} may then be rewritten 
in terms of extended forms as
\eq
\td \;^{s-r-1}\tO_{d+r} + Q \;^{s-r-2}\tO_{d+r+1} 
= d \;^{s-r-1}\OM^r_{d} 
\qquad (\, r=0,\dots,s-1\, )
\ , 
\eqn{ext-eq}
where the $(d+1)$-form on 
the right-hand side cancels the spurious 
$(d+1)$-form which is present on the left-hand side.

Knowing that \equ{ext-eq} is true for $r=0$ (i.e. eq.\equ{bi-desc-0})
and assuming that it holds for $r$, let us prove it for
$r+1$. 
Application of the nilpotent operator  $\td$ to \equ{ext-eq}
yields 
\[
Q\,\td\;^{s-r-2}\tO_{d+r+1} 
= 
d\,\cas\;^{s-r-1}\OM^r_{d} \ .
\]
We now use the $d$-form component of equation \equ{ext-eq},  
which is nothing but the 
bi-descent equation \equ{bi-descent} for $p=d$ 
and a fixed index $r$, to get 
\[
Q\,\td\;^{s-r-2}\tO_{d+r+1} = -d\,Q \;^{s-r-2}\OM^{r+1}_{d} 
= Q\,d \;^{s-r-2}\OM^{r+1}_{d}\ .
\]
Due to the triviality of the $Q$-cohomology, this relation implies the
existence of an extended form $\;^{s-r-3}\tO_{d+r+2}$
and thus leads to equation \equ{ext-eq} with $r+1$.

%%%%%%%%%%%%%%%%%%%%%%%%%%%%%%%%%%%%%%%%%%%%%%%%%

%%%%%%%%%%%%%%%%%%%%%%%%%%%%%%%%%%%%%%%%%%%%%%%%%
\subsection{Proof of Proposition 
\ref{observables in SF form}.4}\label{app-gen-sol}
   
Since we are 
specifically interested in solving, for some 
{\em fixed values of} $d$ {\em and} $D$, 
the bi-descent equations \equ{bi-descent''} 
which correspond to the parallelogram
$\paral$ defined thereafter, 
we have  to restrict the functional space to that of superfield forms 
having SUSY-number $s$ and form degree $p$ constrained by  
\eq
s \leq D-d\ ,\quad p \leq d\ . 
\eqn{cond-trunc}
Thus, we introduce {\it truncated} $q$-{\em superforms} 
(more simply referred to as {\it truncated forms} in the following)
of ghost-number $g$:
\eq
\trO^g_q = \lc \hO^g_q\rc^\trunc \equiv
\dsum{p=q-D+d}{d} 
\;^{q-p}\OM^g_p \; (d\theta)^{q-p} \ .
\eqn{tr-s-forms}
Here, the coefficients $\;^{q-p}\OM^g_p$ are superfield forms.
In the special case where $q+g=D$, 
the truncated form \equ{tr-s-forms}
contains superfield forms of ghost-number $g$ belonging to the 
parallelogram $\paral$.
Depending on the relative values of $g$, $D$ and $d$, the
expansion  \equ{tr-s-forms} may involve terms of negative
SUSY-number or negative form degree, but all of these terms vanish by
virtue of our conventions.

Moreover, we define 
the (nilpotent) {\it truncated differential} $\trd$ 
which has the property of mapping truncated forms to truncated forms:
\eq
\trd \trO =  \lc \hd \trO \rc^\trunc \ ,
\eqn{tr-diff}
We note that the arguments in brackets in the definitions 
\equ{tr-s-forms}\equ{tr-diff}
are superforms. Truncation simply means cutting down all of 
their components which do not satisfy the condition
\equ{cond-trunc}.

The space of truncated forms 
for the fixed pair $(d,D)$ will be denoted by $\tre$.
It is generated by the basic superfield forms $A,\, A_\theta , \, C$
and the action of the 
operators $\cas$, $d$ and $Q$, the latter operator 
giving rise to the expressions 
\[
Q A=\psi, \quad  Q A_\theta=\f,\quad  Q C= c' 
\ .
\] 
The obvious relations 
\eq\ba{c}
 \lc \trO\trFI \rc^\trunc  = \lc \hO\hFI\rc^\trunc  \es
\trd \lc \trO\trFI \rc^\trunc  = \lc (\trd\trO)\trFI
\pm \trO\trd\trFI \rc^\trunc \ ,
\ea\eqn{trunc-form}
which hold for arbitrary truncated superforms $\trO$ and $\trFI$, 
show that the projection from the algebra of superforms 
to the algebra of truncated superforms represents an 
homomorphism. 

In terms of truncated superforms, the bi-descent equations 
for the pair $(d,D)$, i.e. eqs.\equ{bi-descent''}, 
read as 
\eq
\cas\trO^g_{D-g} + \trd\trO^{g+1}_{D-g-1}=0
\qquad( \, g=0,\dots,D \, ) \ .
\eqn{tr-s-descent}
These {\it truncated superdescent equations} define
the cohomology $H(\cas|\trd)$ of $\cas$ modulo $\trd$ 
in the functional space $\tre$. 
This cohomological problem can 
be solved using the algebraic techniques of reference~\cite{barnich}.

Thus, as before, we do not fix the form degree 
and we assume that 
the forms of negative form degree,
ghost-number or SUSY-number vanish. 
The {\em first step} consists of determining the cohomology 
$H(\cas)$ in the functional space $\esup$  
of superfield forms  
introduced in \equ{funct-spaces}
(and subsequently in the functional space $\tre$).
In this respect, it is convenient to consider 
the superfield variables 
$\{\,A$, $A_\theta$, $C$, $\bpsi$, $\bfi$, $K\,\}$ 
 where $\bpsi$, $\bfi$ and $K$ have been 
 defined in eqs.\equ{s-field-red}.
By virtue of the BRST transformations 
\equ{BRS-redef}, the fields 
$ A_\theta$ and $K$ form a BRST doublet and
therefore they are absent from the 
cohomology~\cite{pig-sor,barnich}. 
The remaining fields consist of the 
gauge superfield form $A$ and its ghost $C$, as well as the 
two ``matter superfields'' $\bpsi$ and $\bfi$.
From this fact, we
conclude~\cite{barnich} that the cohomology $H(\cas)$ 
is algebraically
generated by the invariant polynomials 
depending on $C$, the supercurvature ${F_A}$ and  
the matter superfields $\bpsi, \bfi$
-- all of which fields transform covariantly --
as well as 
their covariant exterior derivatives. More precisely, 
the cohomology $H(\cas)$ in the space 
 $\esup$ is generated by the cocycles
\eq
\theta_r(C)\quad (\, r=1,\dots , \mbox{rank }G \, )\quad \mbox{and} 
\quad P^{\rm inv}(F_A ,\, \bpsi,\, \bfi,\, D_A \bpsi ,\,
D_A \bfi )
\ ,
\eqn{s-cohom-AA}
where $\theta_r$ is the cocycle 
\equ{s-cohom} associated to the $r^{\rm th}$ Casimir
operator of the structure group $G$ and where $P^{\rm inv} (\cdots)$ 
is any invariant polynomial of its arguments. 
A straightforward generalization 
of this result from superfield forms to truncated 
superforms yields the following lemma.
%%%%%%%%%%%%%%%%%%%%%%%%%%%%%%%%%%%%%%%%%%%%
\begin{enumerate}\item[]
{\bf Lemma \ref{app}.1} 
The cohomology $H(\cas)$ in the functional space $\tre$ is given by the
truncated forms whose non-vanishing coefficients are polynomials
in the superfield forms given in \equ{s-cohom-AA}.
\end{enumerate}
%%%%%%%%%%%%%%%%%%%%%%%%%%%%%%%%%%%%%%%%%%%%

Let us now determine the cohomology $H(\cas|\trd)$
in the space $\tre$ by 
starting from the 
bottom equation of \equ{tr-s-descent}, i.e. the equation
for $g=D$: $\cas\trO_0^D=0$.
According to lemma \ref{app}.1,
the general non-trivial solution of  the latter 
equation is given by a 
truncated superform $\trO^D_0 = \;^0\OM^D_0$ 
which is a polynomial in the $\theta_r$'s.
However, just as in the case  
of complete superform solutions discussed 
in section \ref{superform sol}, only a linear term 
in $\theta_r (C)$ allows us to work our way up 
to ghost-number zero, i.e. to construct observables. 
Therefore, we assume that $\trO^D_0$ = $\theta_r(C)$ for
some value of $r$. The total degree $D$ will then be odd and given by 
\eq
D=g_r = 2m_r-1\ ,
\eqn{D=g_r}
where $m_r$ is the degree of the $r^{\rm th}$ Casimir operator.

The form $\trO^D_0 = \theta_r(C)$ generates 
a {\em special} solution of the truncated superdescent equations 
\equ{tr-s-descent},
namely the truncation to the
parallelogram $\paral$ of the superforms
$\lc \hat\theta_r \rc^g_{D-g}$ ($\, g=0,\dots,D \,$)
obeying the superdescent equations
\equ{desc-theta}.  
For the top form, this yields
\[
\trO^0_D = \;^{D-d}\Omega^0_d \, (d\theta)^{D-d} = 
\lc \hth^0_D \rc^\trunc\ . 
\]
The {\em general} solution corresponding to the same 
bottom form
$\trO^D_0$ is obtained by adding to it the general solution of the
truncated superdescent equations beginning with $\trO^D_0 = 0$. 
Since we
are only interested in cohomology classes, we shall consider the
slightly more general, though equivalent form
\eq
\trO^D_0 = \cas \trFI^{D-1}_0\ ,
\eqn{initial-Omega}
where $\trFI^{g}_{D-1-g}$ is a truncated  $(D-1-g)$-superform 
of ghost-number $g$.
Solving this problem for a value
of $D$ which is not necessarily equal to $g_r$ as in \equ{D=g_r} will 
give us the general solution of  \equ{tr-s-descent}
corresponding to a bottom form $\trO^D_0$ 
which  is vanishing or trivial in the sense of
eq.\equ{initial-Omega}. 

The procedure is iterative. 
Let us assume that we have arrived, at the stage of ghost-number $g+1$, 
to the trivial solution 
\[
\trO^{g+1}_{D-g-1} = \cas\trFI^{g}_{D-g-1} 
+ \trd \trFI ^{g+1}_{D-g-2}
\]
for some truncated superforms $\trFI^{g}_{D-g-1}$ and
$\trFI ^{g+1}_{D-g-2}$.
By inserting this expression into the descent
equation \equ{tr-s-descent} for ghost-number $g$ and using the nilpotency of $\trd$,
we obtain 
\[
\cas\lp \trO ^g_{D-g} - \trd \trFI ^{g}_{D-g-1} \rp = 0 \ .
\] 
From  lemma \ref{app}.1, it follows that the 
solution of this relation is given by 
\eq
\trO ^{g}_{D-g} = \cas \trFI ^{g-1}_{D-g} 
+ \trd \trFI ^{g}_{D-g-1}
+ \trHH ^{g} _{D-g}  \ ,
\eqn{sol-level-g}
where the truncated superform $\trHH^{g}_{D-g}$ 
belongs to the cohomology $H(\cas)$ as given by lemma 
\ref{app}.1,
if there exists a representative of the latter 
with the right ghost-number and form degree.
The supercocycle $\trHH ^{g} _{D-g}$
has to satisfy a consistency condition 
ensuring that the next descent equation 
in \equ{tr-s-descent} is integrable. 
Indeed, let us substitute \equ{sol-level-g}
into the descent equation for ghost-number $g-1$, thus obtaining
\eq
\cas\lp \trO ^{g-1} _{D-g+1} - \trd \trFI ^{g-1} _{D-g} \rp 
+ \trd \trHH ^g_{D-g} = 0\ .
\eqn{eq-level(g-1)}
In order for this equation to admit a solution $\trO ^{g-1} _{D-g+1}$, 
there must exist a truncated superform $\trH ^{g-1} _{D-g+1}$ such that
\eq
\cas \trH ^{g-1} _{D-g+1}  +  \trd \trHH ^g _{D-g} = 0 \ .
\eqn{consist-HH} 
Then, the solution of \equ{eq-level(g-1)} is given by 
\eq
\trO ^{g-1} _{D-g+1} 
= \cas \trFI ^{g-2} _{D-g+1} 
+ \trd \trFI ^{g-1} _{D-g}
+ \trH ^{g-1} _{D-g+1} + \trHH ^{g-1} _{D-g+1}  
\ ,
\eqn{sol-level(g-1)}
where $\trHH ^{g-1} _{D-g+1}$ is again an element of
$H (\cas)$ 
which, in turn, has to obey a consistency condition like
\equ{consist-HH}.

Discarding for the moment all of the supercocycles $\trHH$ which 
have appeared or may still appear
during this process, we finally arrive, at
ghost-number zero, to the trivial solution
\[
\trO^{0}_{D} =  \trd \trFI^{0}_{D-1}  \ ,
\]
which, according to 
definition \equ{tr-s-forms}, reads explicitly as
\eq
^{D-d}\OM^0_d = d \;^{D-d}\Phi^0_{d-1} + Q \;^{D-d-1}\Phi^0_{d}  \ .
\eqn{sol-level-0}
 By virtue of  equation \equ{obs-d-s},
this solution corresponds to a vanishing observable: 
$^{D-d+1} {\Delta}_{(d)} \equiv \int_{M_d} Q \; ^{D-d}\OM^0_d =0$. 

Let us now go back to one of the steps where a cohomological term 
$\trHH$ appears. 
 Since this term belongs to 
 the cohomology $H(\cas)$, it is
a polynomial in the cocycles \equ{s-cohom-AA} -- again to be taken as 
(at most) linear in the $\theta_r (C)$. Thus, we 
consider a generic term of one of the two following forms, 
which generalizes the
superform expression \equ{s-d-cohom}: 
\eq
\trHH^{0}_{D} = 
 \trP_{D}^0(F_A ,\, \bpsi,\, \bfi ,\, D_A \bpsi ,\, D_A \bfi ) 
\eqn{HH-0}
or 
\eq
\trHH^{g_{r_1}}_{D-g_{r_1}} 
= \theta_{r_1}(C) \ \trP_{D-g_{r_1}}^0
(F_A ,\, \bpsi,\, \bfi ,\, D_A \bpsi ,\, D_A \bfi ) \ .
\eqn{HH-G} 
Here, $\theta_{r_1} (C)$ denotes 
the cocycle \equ{s-cohom} of ghost-number $g_{r_1}$, while 
the truncated superforms $\trP_D^0$ and 
$\trP_{D-g_{r_1}}^0$ are invariant polynomials of their arguments.

Let us begin with the solution \equ{HH-0} which may
be encountered in the last step of the 
process described above, namely at ghost-number zero. 
In this case, 
the coefficient $\;^{D-d}\ZZ^0_d$ of the truncated form 
$\trHH^0_D = \;^{D-d}\ZZ^0_d \; ( d\theta )^{D-d}$
is a BRST invariant polynomial in the superfield
forms $F_A$, $\bpsi$, $\bfi$ and their covariant derivatives as in
equation \equ{s-cohom-AA}.
By virtue of \equ{sol-level-g}  
with $g=0$, the expression 
$\trO^0_D = \trHH^0_D$ solves the truncated superdescent equations 
\equ{tr-s-descent}. 
This solution is 
cohomologically non-trivial if 
\eq
\;^{D-d}\ZZ^0_d \not= d \;^{D-d} \Phi_{d-1}^0 + Q  \;^{D-d-1}\Phi_{d}^0 
\ .
\eqn{OM-eq-triv}
This result 
yields the second class of solutions
announced in proposition \ref{observables in SF form}.4, 
i.e. the one given in equation \equ{gen-sol-2}.
%%%%%%%%%%%%%%%%%%%%%%%%%%%%%%%%%%%%%%%%%%%%%%
 
Next, we turn 
to the case given by the solution \equ{HH-G},
which case may be encountered at a ghost-number $g_{r_1} >0$.
We now have to solve the consistency condition \equ{consist-HH}.
We recall that the cocycle $\theta_{r_1}$ 
generates a set of (complete) superforms 
$\lc \hat\theta_{r_1} \rc^{g_{r_1}-p}_{p}$ ($\, p=0,\dots,g_{r_1} \,$)
obeying the superdescent equations \equ{desc-theta}.
Substituting the expression \equ{HH-G} into
\equ{consist-HH} and using the superdescent equation \equ{desc-theta} 
for the ghost cocycle 
$\lc \hat\theta_{r_1} \rc^{g_{r_1}-1} _1$, we obtain 
the following relation with the help
of the properties \equ{trunc-form}: 
\eq
\cas \trH ^{g_{r_1}-1} _{D-g_{r_1}+1}
\, - \, 
\lc \lp \cas \lc \hat\theta_{r_1} \rc 
^{g_{r_1}-1}_1 \rp \trP^0_{D-g_{r_1}} \rc^\trunc
\, + \, 
(-1)^{g_{r_1}} \; \theta _{r_1} (C) \; \trd \trP^0_{D- g_{r_1}} 
= 0 \ . 
\eqn{xzzs}
Here, the exponent `tr' of the second term means
truncation  according to the definition \equ{tr-s-forms}.
The last term 
in \equ{xzzs} is a non-trivial $\cas$-cohomology class,  
whereas the first two terms are $\cas$-exact.
 Therefore, both 
expressions must vanish separately. This implies the following 
{\it consistency condition} for $\trP$:
\eq
\trd \trP^0_{D-g_{r_1}} = 0 \ .
\eqn{P-consistency}
Assuming this relation to hold, condition \equ{xzzs} can now 
be solved by 
\eq
\trH ^{g_{r_1}-1}_{D-g_{r_1}+1} = \lc  
\lc \hat\theta_{r_1} \rc 
_1^{g_{r_1}-1} \trP^0_{D-g_{r_1}} \rc^\trunc  \ ,
\eqn{sol-H-first}
where we discarded possible $\cas$-exact terms as well as 
terms belonging to $H(\cas)$ which, for their part, 
would generate further solutions.
As a matter of fact, the solution \equ{sol-H-first} is the first 
of a chain of truncated supercocycles
\eq
\trH^g_{D-g} 
\equiv \lc \hthe{1}^{g}_{g_{r_1}-g} \trP^0_{D-g_{r_1}} \rc^\trunc 
\quad (\, g=0,\dots,g_{r_1}-1  \, ) \ , 
\eqn{sol-H-gen}
which obey the following truncated superdescent 
equations by virtue of
eqs.\equ{desc-theta} and \equ{P-consistency}: 
\eq
\cas \trH^g_{D-g} + \trd \trH^{g+1}_{D-g-1} =0 \quad ( \, g 
=0,\dots,g_{r_1}-1 \, )\ .
\eqn{H-desc-eq}

We still have to solve condition \equ{P-consistency}. 
This requires the determination of the 
cohomology $H(\trd)$ in the space $\tre$, the
result being expressed by the following lemma:
%%%%%%%%%%%
\begin{enumerate}\item[] 
{\bf Lemma  \ref{app}.2} 
The cohomology $H (\trd)$ in the space $\tre$ is given by the truncated
superforms of ghost-number $0$ and total degree $D$:
\eq
\trO^0_D = 
\;^{D-d}\OM^0_d \; ( d\theta )^{D-d} 
\ .
\eqn{cohom-trd} 
Here, 
$\;^{D-d}\OM^0_d$ is a BRST invariant polynomial in the superfield
forms $F_A$, $\bpsi$, $\bfi$ and their covariant derivatives as in
eq.\equ{s-cohom-AA}, of degree $d$ and ghost-number $0$,
but subject to the non-triviality condition
\[
\;^{D-d}\OM^0_d \not= d \;^{D-d} \Phi_{d-1}^0 + Q  \;^{D-d-1}
\Phi_{d}^0 \,
 \ ,
\]
where the superfield forms $^{D-d}\Phi_{d-1}^0$ and 
$\;^{D-d-1}\Phi_{d}^0$
 are the components of a truncated superform
belonging to $\tre$.
In particular,    
the cohomology $H(\trd)$ is trivial for truncated superforms of
degree strictly      
smaller than $D$.
\end{enumerate}
%%%%%%%%%%%%%%%%%%%%%%%%%%
\noindent {\bf Proof:} 
In this proof, we do not specify the ghost-number which is
irrelevant for the present discussion.
We have to solve the equation $\trd\trO_q = 0$
for the truncated form \equ{tr-s-forms}.
Let us begin with the generic case, i.e. $q<D$.
The condition $\trd\trO_q = 0$ can then  
be written as a set of equations, one for each
form degree: 
\eq
Q \;^{q-p-1}\OM_{p+1} + d \:^{q-p}\OM_{p} = 0
\quad ( \, p=q-D+d,\dots,d-1 \, ) \ . 
\eqn{tr2}
The first of these equations, namely the one for $p=q-D+d$, may be solved 
by using proposition \ref{observables in SF form}.1 
(see \equ{lemma1a}-\equ{L1b'}), which yields
\eq\ba{l}
\;^{D-d}\OM_{q-D+d} = Q \;^{D-d-1}\Phi_{q-D+d} 
+ d \;^{D-d}\Phi_{q-D+d-1}  \es
\;^{D-d-1}\OM_{q-D+d+1} = Q \;^{D-d-2}\Phi_{q-D+d+1} 
+ d \;^{D-d-1}\Phi_{q-D+d}  \ .
\ea\eqn{trr1}
Substituting this result into the second equation of the set
\equ{tr2}, we get
\[
Q\lp \;^{D-d-2}\OM_{q-D+d+2} -d \;^{D-d-2}\Phi_{q-D+d+1}  \rp = 0\ ,
\]
whose solution, due to the triviality of $H(Q)$, is given by 
\eq
\;^{D-d-2}\OM_{q-D+d+2} = Q \;^{D-d-3}\Phi_{q-D+d+2} 
+ d \;^{D-d-2}\Phi_{q-D+d+1}  \ .
\eqn{trr2}
The procedure continues along these line until
the solution of the last equation of the set \equ{tr2}:
\eq
\;^{q-d}\OM_{d} = Q \;^{q-d-1}\Phi_{d} 
+ d \;^{q-d}\Phi_{d-1}  \ .
\eqn{trrr2}
By combining these results, we conclude that, for $q<D$, we have  
\eq
\trO_q= \trd\trFI_{q-1}\ , \qquad  
{\rm with} \ \; 
\trFI_{q-1} = \dsum{p=q-D+d-1}{d} \;^{q-1-p}\Phi_p
\; (d\theta )^{q-1-p} \ ,
\eqn{tr3}
i.e. the general solution is trivial.
Thus, we are left with the case $q=D$, 
i.e. the cocycle condition $\trd\trO_D = 0$, 
which is identically satisfied, whence the result
\equ{cohom-trd}. 
\hfill \cqfd
%%%%%%%%%%%%%%%%%%%%%%%%%%%%%%%%%%%%%%%%%%%%%%%%%%%%%
%%%%%%%%%%%%%%%%%%%%%%%%%%%%%%%%%%%%%%%%%%%%%%%%%%%%%

Let us now solve equation \equ{P-consistency} with the help of this
lemma. 
Since $D-g_{r_1}<D$, it follows from lemma \ref{app}.2  
that the solution $\trP^0_{D-g_{r_1}}$ 
of \equ{P-consistency} is a $\trd$-variation, i.e.
$\trP^0_{D-g_{r_1}} = \trd \trR^0_{D-g_{r_1}-1}$. 
Thereby, equation \equ{sol-H-gen} for $g=0$ can be written as
\eq
\trH^{0}_{D} 
= \lc \hthe{1}^0_{g_{r_1}}  \trd \trR^0_{D-g_{r_1}-1} \rc^\trunc \ .
\eqn{HH-G-1}
Since $\trP^0_{D-g_{r_1}}$ is $\cas$-invariant, the polynomial 
$\trR^0_{D-g_{r_1}-1}$ again 
has to be a solution of a system of
truncated superdescent equations in the functional space $\tre$:
\eq\ba{l}
\cas \trR^{g}_{D-g_{r_1}-g-1} + \trd 
\trR^{g+1}_{D-g_{r_1}-g-2} = 0 \qquad
(g= 0,\dots , D-g_{r_1}-1) \ .
\ea\eqn{tr-desc-2}
These are solved in the same way
as we did in the first step above\footnote{However, there is a
difference here:  since the total degree is less than  $D$ and the
truncation is made relative to the pair $(d,D)$, the parallelogram
is replaced by a pentagram  defined by the lines 
\[
p=0,\ p=d,\ s=0,\ s=D-d, \ g=0 \ .
\]
}. 
The non-trivial solution of the bottom equation 
is a supercocycle 
\eq
\trR^{g_{r_2}}_{0} = \theta_{r_2}(C)
\qquad {\rm with} \ \ g_{r_2} \equiv D-g_{r_1}-1 \ ,
\eqn{bottom-2}
if any such exists with this
ghost-number. Otherwise $\trR^{g_{r_2}}_{0}$ $=$ 0 and one has to climb
up equations \equ{tr-desc-2} until meeting a non-trivial cohomology
-- as we did in the first step -- and then continue, 
starting from this cohomology. But let us
consider the case where \equ{bottom-2} holds. Then, 
we get the following result by using \equ{desc-theta} and by 
discarding possible new cohomology 
that may appear in the process of climbing up:  
\[
\trR_{D-g_{r_1}-1}^{0} 
= \lc\lc\hat\theta_{r_2}\rc^0_{g_{r_2}} \rc^\trunc \ .
\]
Substitution of this expression into \equ{HH-G-1}
and application of the rules \equ{tr-diff}-\equ{trunc-form} then yields 
\eq
\trH^{0}_{D} 
= \lc \hthe{1}^0_{g_{r_1}} \hd \hthe{2}^0_{g_{r_2}} \rc^\trunc
= \lc \hthe{1}^0_{g_{r_1}} f_{r_2}(\hF) \rc^\trunc \ ,
\qquad\mbox{with}\quad g_{r_1}+ g_{r_2} +1=D\ ,
\eqn{HH-G-2}
where we have used the last of equations \equ{desc-theta}.

Going back to the previous step where new cohomology might have been
encountered, we may repeat the whole argument, producing in this way
solutions involving more and more factors 
$\hd\hth^0_{g_r}$ = $f_{r}(\hF)$. Thus, the 
 generic solution of the truncated superdescent 
equations reads as 
$\trO^0_D = \check{H}^0_D \equiv \;^{D-d}H^0_d \; ( d\theta )^{D-d}$
where 
\eq
\trH^{0}_{D} = \lc \hthe{1}^0_{g_{r_1}} f_{r_2}(\hF) \cdots
f_{r_L}(\hF)  \rc^\trunc\ ,\qquad \mbox{with}\quad
\dsum{k=1}{L} g_{r_k} + L -1=D\quad     (L\ge1) \ .
\eqn{HH-G-L}
This conclusion is precisely the result 
\equ{gen-sol-1} stated in proposition 
\ref{observables in SF form}.4. 

%%%%%%%%%%%%%%%%%%%%%%%%%%%%%%%%%%%%%%
%%%%%%%%%%%%%%%%%%%%%%%%%%%%%%%%%
%%%%%%%%%%%%%%%%%%%%%%%%%%%%%%%%%%%%%%%%%%%%%%%%%

%%%%%%%%%%%%%%%%%%%%%%%%%%%%%%%%%%%%%%%%%%%%%%%%%%
\end{document}